\documentclass{FBSsuppl}
\usepackage{citesort,psfig,epic,eepic,amsfonts,amssymb,amsmath,rotating,blockeqart}
\renewcommand{\theequation}{\thesection.\arabic{equation}\alph{subequation}}
\newcounter{subequation}
\newcommand{\Eq}{Eq.\@~{}}
\newcommand{\Eqs}{Eqs.\@~{}}
\newcommand{\Fig}{Fig.\@~{}}
\newcommand{\Figs}{Figs.\@~{}}
\newcommand{\Ref}{ref.\@~{}}
\newcommand{\Refs}{refs.\@~{}}

\newcommand{\Diso}{$\Delta$-Isobar{}}
\newcommand{\diso}{$\Delta$-isobar{}}
\newcommand{\rmDelta}{\Delta}%

\newcommand{\hethree}{${}^3$He}

\newcommand{\ltau}{\mbox{\Large $\tau$}}
\newcommand{\lz}{\mbox{$Z$}}

\renewcommand {\vec}[1]{{\bf  #1 }}

\newcommand {\bra}[1]{\langle #1 \vert}
\newcommand {\ket}[1]{\vert #1 \rangle}
\newcommand {\hatbf}[1]{\hat{ \bf  #1} }
\newcommand {\braket}[2]{\langle #1 \vert #2 \rangle}
\newcommand {\oot}{\frac{1}{2}}
\newcommand {\thot}{\frac{3}{2}}
\newcommand {\fcal}[1]{{\cal #1}}
\renewcommand {\vec}[1]{{\bf  #1 }}
\newlength{\mylen}

\title{Trinucleon Photo Reactions with ${\Delta}$-Isobar Excitation:\\ Radiative Nucleon-Deuteron Capture and \\
   Two-Body Photo Disintegration of the Three-Nucleon Bound State}

\author{L.~P.~Yuan\instnr{1}, 
	   K.~Chmielewski\instnr{1},
	   M.~Oelsner\instnr{1}, P.\@ U.\@ Sauer\instnr{1},
	   A.~C.~Fonseca\instnr{2}, 
	   and \\ J.~Adam~Jr.\instnr{3} }
\instlist{Institut f\"ur Theoretische Physik, Universit\"at Hannover,
	     Appelstra{\ss}e 2, \mbox{D-30167} Hannover, Germany
	   \and Centro F\'{\i}sica Nuclear da Universidade de
	   Lisboa,
	  Av.\@ Prof.\@ Gama Pinto, N$^\circ$2, P-1649-003 Lisboa, Portugal
	   \and Nuclear Physics Institute, Academy of Sciences of
	   Czech Republic, CZ-25068, \v{R}e\v{z}, Czech Republic }

\begin{document}

\maketitle

\begin{abstract}
 Radiative nucleon-deuteron capture and  photo disintegration of the three-nucleon bound state
  with two-body final states are described. The description
uses nucleon degrees of freedom extended to include the excitation of
 a single nucleon to a \diso. The baryonic interaction and the electromagnetic current couple nucleonic states and states with a \diso. Exact solutions of three-particle scattering equations are employed for the initial or final states of the reactions. The current has one-baryon and two-baryon contributions. The role of the \diso~in the description of the considered photo reactions is discussed and found to be moderate. The spin observables $A_{yy}$ and $T_{20}$ at $90^{\circ}$ lab scattering angle can be calculated model-independently from the $E1$ Siegert term in the long-wavelength limit.
\end{abstract}

\newpage
\newcommand\themysection{\thesection}

\section{Introduction}
\setcounter{equation}{0}
\label{sec:intro}

Trinucleon photo reactions are theoretically described. Only
reactions with two-body initial or final states are discussed, not three-body photo disintegration of the three-nucleon bound state. The considered reaction energies are
well above three-nucleon break-up, but remain otherwise low, i.e., below 50 MeV in the three-nucleon c.m. system.

The present calculation of trinucleon photo reactions extends the description of elastic and inelastic nucleon-deuteron scattering given by some of the authors in \Refs \cite{nemoto:98a,nemoto:98b,nemoto:98c,chmielewski:98a99b}. Nucleon-deuteron scattering is studied in general \cite{gloeckle:96a}, in order to search for unambiguous signals of a three-nucleon force. Indeed, discrepancies between theoretical predictions, derived from two-nucleon potentials, and experimental data got uncovered \cite{witala:98a,nemoto:98c}. In the same spirit trinucleon photo reactions are studied, in order to find clear indications for the working of a three-nucleon force and of corresponding two- and three-nucleon electromagnetic (e.m.) exchange currents.

The theoretical description of trinucleon photo reactions, given in this paper, is realistic in the following sense: A realistic nuclear interaction is employed when solving for the hadronic states involved, i.e., for the trinucleon bound state and for the nucleon-deuteron scattering states. The employed nuclear interaction is a hermitian coupled-channel two-baryon potential between nucleons which allows for the excitation of a single nucleon to a $\rmDelta$-isobar. The $\rmDelta$-isobar mediates an effective three-nucleon force. However, the Coulomb interaction between the two protons is not taken into account. Thus, the calculations of this paper are done for photo reactions on ${}^3$H, though most experimental data will refer to photo reactions on \hethree. The nuclear current is built-up from one-baryon and two-baryon contributions. The current contributions contain, for reasons of consistency, pieces with transitions to one-baryon and two-baryon states with a single $\rmDelta$-isobar. The description of the considered reactions is comparatively transparent: As long as only a few multipoles of the current determine the observables, only few total angular momenta are picked out from the nucleon-deuteron scattering states.

The calculation of trinucleon elastic and inelastic nucleon-deuteron scattering, given in \Refs \cite{nemoto:98a,nemoto:98b,nemoto:98c,chmielewski:98a99b}, is based on a separable expansion of the two-baryon transition matrix; the same separable expansion is used here for the description of photo reactions. In the separable treatment of the hadronic interaction, our calculation corresponds to the work of \Refs \cite{fonseca:93a,fonseca:00a}; however, it is more complete with respect to the employed current. Our calculation attempts to be as realistic as the description given in \Refs \cite{anklin:98a,golak:00a}, which does not require a separable expansion of the hadronic interaction. In contrast, the addition of the \diso~degree of freedom gives our calculation a new theoretical dimension. The calculation will attempt to isolate the effects arising from the $\rmDelta$-isobar. Preliminary results were presented in \Ref \cite{yuan:00a}.

Sect. 2  describes the calculational apparatus. Sect. 3  shows and discusses results. Sect. 4  is a summary with conclusions.

\section{Calculation Procedure}
\subsection{Basic Elements of Calculation}
The basic building block for the description of photo reactions is the e.m. hamiltonian
\begin{equation}
H^{e.m.}_I = \frac{1}{c} \int d^3 x \; J^{\mu} (x) A_{\mu}(x) |_{x_0 = 0} \hspace{0.0cm} , \label{hamilton}
\end{equation}
which couples baryonic states to the photon ($\gamma$). The operators in \Eq (\ref{hamilton}) depend on space-time $x$, but are to be used as Schr\"odinger operators at time $ x_0 = 0 \hspace{0.1cm}; \; A_{\mu}(x)$ is the photon field operator, which we parametrize in the form
\begin{gather}
\begin{align}
 A_{\mu}(x) = &\frac{(4 \pi)^{1/2} \hbar c }{(2 \pi ) ^{3/2}}\int \frac{d^3 k_{\gamma}}{\sqrt {2 k_{\gamma 0} c} } \sum ^3_{\lambda = 0}  \nonumber \\
&   
\big [ a_{\lambda} ( \vec{k}_{\gamma} ) \epsilon _{\mu} ( \vec{k}_{\gamma} \lambda) e^ {-\frac{i }{\hbar}  k_{\gamma} x }  + a_{\lambda}^{\dagger} ( \vec{k}_{\gamma} ) \epsilon _{\mu} ^{*} ( \vec{k}_{\gamma} \lambda) e^ {\frac{i }{\hbar}  k_{\gamma} x }  \big ] |  _{k_{\gamma 0 } = | \vec{k} _{\gamma} | }  \hspace{0.05cm}.
\end{align}
\end{gather}
The polarizations $\lambda = 1,2$ correspond to those of a real transverse photon;  $ \epsilon _{\mu} ( \vec{k}_{\gamma} \lambda) $ are the polarization  vectors, constrained by $ k_{\gamma} ^{\mu }\epsilon _{\mu} ( \vec{k}_{\gamma} \lambda) = 0 $. A single-photon state of definite momentum $\vec{k}_{\gamma} $ and polarization $\lambda $, $\delta $-function normalized, is $ |\vec{k}_{\gamma} \lambda \rangle = a ^{\dagger} _{\lambda} ( \vec{k}_{\gamma} )|0 \rangle $.

\begin{figure}[h]
\psfig{file=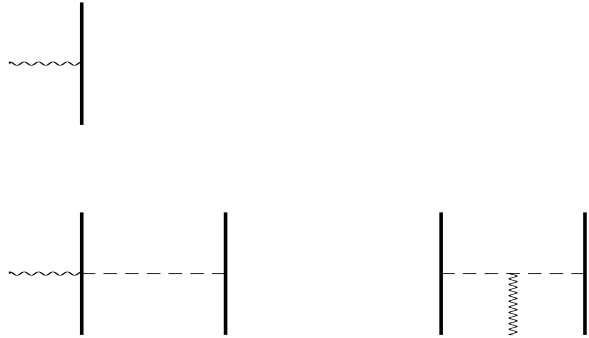,width=380pt,height=150pt}
%\parbox[t]{\bild}{\psfig{file=onecur.eps,width=\bild,height=40pt }}
\caption{\sloppy One- and two-baryon processes contained in the used e.m. current. The dashed horizontal line indicates instantaneous meson exchange. Diagonal  $\pi$- and $\rho$-exchanges are taken into account as well as the nondiagonal  $\rho\pi\gamma$ and $\omega\pi\gamma$ contributions. In this figure only purely nucleonic processes are depicted.}
\label{fig:onecur} 
\end{figure}

\begin{figure}[htb]
\psfig{file=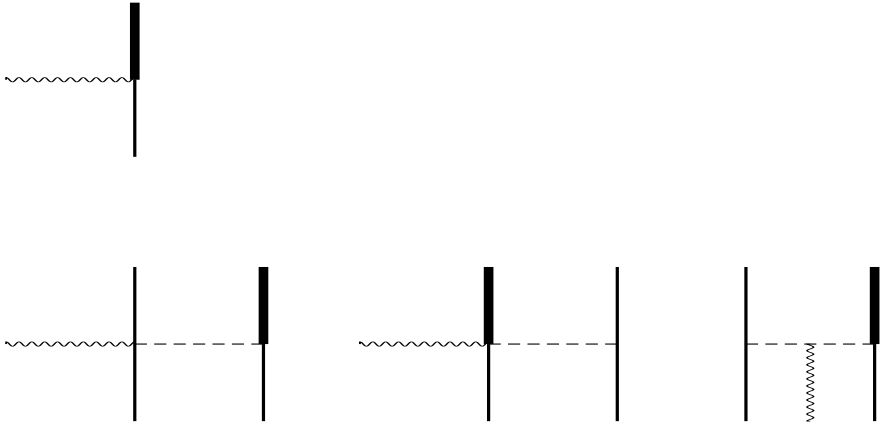,width=340pt,height=130pt}
\caption{One- and two-baryon processes contained in the used e.m. current. In this figure processes are depicted in which one nucleon is turned into a \diso. The hermitian adjoint processes are taken into account, but are not diagrammatically shown.}
%In contrast to the processes of \Fig \ref{fig:onecur} only diagonal $\pi$-and $\rho$-exchanges are taken into account.}
\label{fig:twocur} 
\end{figure}

\begin{figure}[h]
\psfig{file=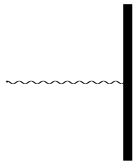,width=480pt,height=160pt}
\vspace{-80pt}
\caption{One-baryon process contained in the used e.m. current. It is the only process between \diso~states which is considered.}
\label{fig:delcur} 
\end{figure}

The e.m. current operator $  J^{\mu} (x) $ acts in the baryonic Hilbert space which we take to have two sectors, i.e., a purely nucleonic one and one in which one nucleon ($N$) is turned into a $ \rmDelta $-isobar. The two Hilbert sectors are displayed in \Fig \ref{hilbertspace} for a three-baryon system as ${}^3$He. We use the current operator in its Fourier-transformed form, i.e.,
\begin{equation}
  J^{\mu} ( \vec{Q}) =  \int d^3 x \; e^  {\frac{i }{\hbar} \vec{Q} \cdot \vec{x} }  J^{\mu} (x) |_{ x_0 = 0} \hspace{0.0cm},
\end{equation}
 and employ --- specializing to a three-baryon system --- a momentum-space representation, based on the Jacobi momenta ( $ \vec{p} \vec{q} \vec{K} $ ) of  three particles in the definition of ref. \cite{nemoto:98a}, i.e.,
\begin{equation}
 \langle \vec{K} ' \vec{p} ' \vec{q} ' |  J^{\mu} ( \vec{Q}) | \vec{K}  \vec{p}  \vec{q} \rangle = \delta ( \vec{K} ' - \vec{Q} - \vec{K} )   \langle  \vec{p} ' \vec{q} ' |  j^{\mu} ( \vec{Q} , \vec{K} ' + \vec{K} ) |   \vec{p}  \vec{q} \rangle \hspace{0.0cm} .
\label{currentj}
\end{equation}
In \Eq (\ref{currentj}) $\vec{Q}$ is the three-momentum transfer by the photon; it will take on particular values depending on the considered reaction; in e.m. reactions with real photons it is given by the photon momentum $ \vec{k} _ {\gamma}$. A total-momentum conserving $ \delta$-function is split off;  the remaining current operator $   j^{\mu} ( \vec{Q} ,\vec{K} '+ \vec{K} )  $ only acts on the internal momenta of the three-baryon system with a parametric dependence on the combination $\vec{K} '+ \vec{K} $ of total momenta. Since all meson degrees of freedom are frozen, the operator has one-baryon and many-baryon pieces. Besides the standard nucleonic-current part there are additional parts involving the $\rmDelta $-isobar. We take one-baryon and two-baryon contributions into account, shown in Figs.{} \ref{fig:onecur} - \ref{fig:delcur}. The horizontal lines in the diagrams  indicate that the meson exchanges are instantaneous. The exchanged mesons
%\enlargethispage*{220cm}
\newlength{\bild} 
\setlength{\bild}{0.895\linewidth}
\begin{figure}[htb]
%\parbox[t]{\bild}{\psfig{file=onecur.eps,width=\bild}}
\parbox[t]{\bild}{\psfig{file=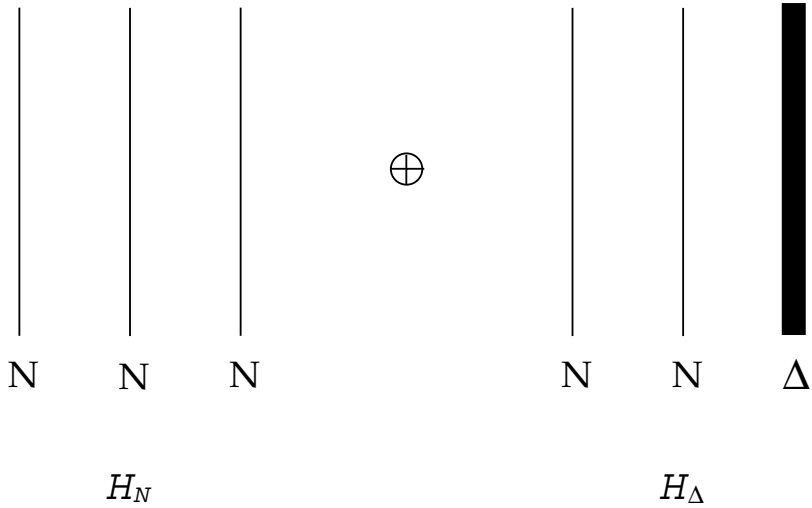,width=\bild}}
\vspace{0pt}\caption{\sloppy Hilbert space considered. It consists of a purely nucleonic sector $\mathcal{H}_N$ and a sector $\mathcal{H}_{\Delta} $ in which one nucleon is turned into a \diso {}, indicated by a thick vertical line.}
\label{hilbertspace} 
\end{figure}
\vspace{-40pt}
 are the pion ($\pi$) and the rho ($\rho$). The nondiagonal $\rho\pi\gamma$ and $\omega\pi\gamma$ contributions, $\omega$ denoting the isoscalar vector meson omega, are also taken into account for the purely nucleonic current of \Fig \ref{fig:onecur}. The current of \Fig \ref{fig:twocur} couples purely nucleonic states with states containing one \diso; the hermitian adjoint pieces are not diagrammatically shown; the nondiagonal $\rho\pi\gamma$ and $\omega\pi\gamma$ contributions to the two-baryon current involving one \diso~are not considered. Among the contributions between \diso~states only the one of one-baryon nature is kept as shown in \Fig \ref{fig:delcur}. The current is derived by the extended $S$-matrix method of refs. \cite{strueve:87a,adam:89a,adam:91a,henning:92a}; it satisfies current conservation with the corresponding $\pi$- and $\rho$-exchanges in the coupled-channel two-baryon interaction $H_I$ of one-boson exchange nature. The current is systematically expanded up to first order in $p/m_N$, $\vec p$ being a characteristic baryon momentum and $m_N$ the nucleonic rest mass. \\
%\clearpage

\vspace{40pt}
\vspace{-20pt}
%\vspace{11cm}
In the perturbative spirit for the evolution of photo processes, the e.m. interaction $ H^{e.m.} _I $ acts only once, whereas the hadronic interaction $ H_I $ has exactly to be taken into account up to all orders. We use hadronic channel states, seen  in the initial and final states $ | i \vec{P} _i \rangle $ and $ | f \vec{P} _f \rangle $ of the photo reactions, in the form
\blockeqn
\begin{align}
  | \Phi _B \vec{p}_B \rangle & = | B \rangle | \vec{p}_B \rangle \hspace{0.0cm} ,  \\ 
  | \Phi _ {\alpha} ( \vec{q} ) \nu _ {\alpha} (Nd) \vec{K} \rangle & =  | \phi
    _ {\alpha} ( \vec{q} ) \nu _ {\alpha} (Nd) \rangle | \vec{K} \rangle \hspace{0.0cm} ,
\end{align} \reseteqn
with the energies
\blockeqn \onelabel{kineen} 
\begin{align} 
E_B ( \vec{p}_B ) &= m_B c ^2 + \frac{\vec{p}_B ^2}{6 m _N} \hspace{0.0cm} , \\ E _{Nd} ( \vec{q} \vec{K} ) &= m_N c^2 + m_d c^2 + \frac{\vec{q} ^2}{4 m_N /3} + \frac{\vec{K} ^2}{6 m _N} \hspace{0.0cm} , \label{enerkine}
\end{align}
\reseteqn
$m_N, m_d$ and $m_B$ being the rest masses of nucleon, deuteron and trinucleon bound state. The internal trinucleon bound state is $ | B \rangle $, which is normalized to 1. The product nucleon-deuteron channel state in the three-nucleon c.m. frame is $| \phi _ {\alpha} ( \vec{q} ) \nu _  {\alpha} (Nd) \rangle $  in the notation of ref. \cite{nemoto:98a}, $ \nu _{\alpha} $ denoting all discrete quantum numbers. In both cases the c.m. motion is explicitly added to the internal motion; in the three-nucleon channel with a photon, the total momentum ${\vec P}$ is different from the total momentum ${\vec K} = {\vec p}_B$ of the three nucleons, bound in the trinucleon bound state $| B \rangle $. 

The matrix elements of the e.m. interaction require fully correlated hadronic states, i.e.,
\blockeqn
\begin{align} 
 | \Phi _B \vec{p}_B \rangle &=  \pm i0G\big(E_B ( \vec{p}_B ) \pm i0 \big) | B \rangle | \vec{p}_B \rangle \hspace{0.0cm}  ,   \\ 
| \Psi ^{ (\pm )} _ {\alpha} ( \vec{q} ) \nu _  {\alpha} (Nd) \vec{K} \rangle &=  \pm i0G\big(E _{Nd} ( \vec{q} \vec{K} ) \pm i0\big)\big(1+P\big)/\sqrt{3}| \Phi _ {\alpha} ( \vec{q} ) \nu _  {\alpha} (Nd) \vec{K} \rangle\hspace{0.0cm} , \label{eq:scatsta}
\end{align}
\reseteqn
with the full resolvent
\begin{gather}
G(Z) = \frac{1}{Z-H_0 - H_I}\hspace{0.0cm}  ;
\end{gather}
the free Hamiltonian $ H_0 $ contains the motion of the center of mass; $P$ symmetrizes the nucleon-deuteron product state; the individual kinetic energy operators are of nonrelativistic form; they yield the eigenvalues of \Eqs (\ref{kineen}). The hadronic state \eqref{eq:scatsta} is normalized to the delta
function without additional normalization factors. Since the hadronic interaction Hamiltonian $ H_I $ acts on relative coordinates only, the full resolvent reproduces the bound state $ | B \rangle $ and correlates the nucleon-deuteron scattering state only in its internal part, i.e.,
\begin{equation}
| \Psi ^{ (\pm )} _ {\alpha} ( \vec{q} ) \nu _  {\alpha} (Nd) \vec{K} \rangle = | \psi ^{ (\pm )} _ {\alpha} ( \vec{q} ) \nu _  {\alpha} (Nd)\rangle\ |  \vec{K} \rangle \hspace{0.0cm} .
\end{equation}
\newpage
\noindent 
\subsection{${S}$-Matrix for Trinucleon Photo Reactions}
\label{sec:smatrix}
\begin{figure}[htb]
\begin{center}
\psfig{file=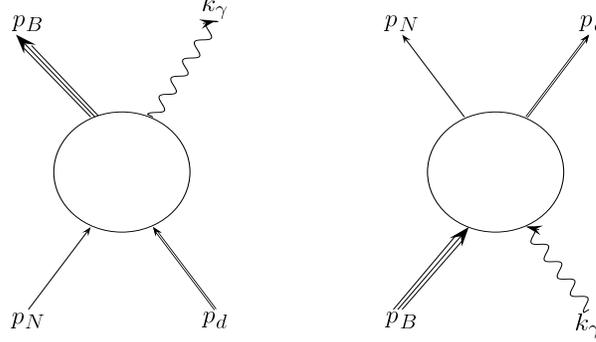,width=350pt,height=105pt,angle=-90}
\vspace{10pt}\caption{Schematic description of the considered photo reactions of the deuteron. The lines for the two-baryon and three-baryon particles are drawn in a special form to indicate their compositeness.}
\label{fig:reaction}
\vspace{-20pt}
\end{center}
\end{figure}

The kinematics of the considered processes is shown in \Fig \ref{fig:reaction}. The figure also defines the employed notation for the individual particle momenta; $p_N,p_d,p_B$ and $ k_{\gamma} $ are on-mass-shell four-momenta.
 The corresponding particle energies are derived from the zero components of those momenta, i.e., $E_N ( \vec{p} _N) = p_{N0}c,\; E_d( \vec{p} _d)= p_{d0}c  $ and $
E_B( \vec{p} _B ) = p_{B0}c  $; they are relativistic ones in contrast to those of the nonrelativistic model calculation in \Eqs (\ref{kineen}) for which the same symbols are used; we shall make sure in the following that no confusion arises.

The $ S$-matrix and the differential cross sections take the forms given in the following subsections. 
 \subsubsection{ Radiative Capture in Nucleon-Deuteron Scattering}
\label{sec:rad}
We give various alternative forms for the $S$-matrix elements:
\blockeqn \onelabel{smatrix}
\begin{align} 
 \langle f \vec{P} _f | S^{rc} | i \vec{P} _i \rangle & =  (-i)(2 \pi \hbar ) ^{4}  \delta \big( k _{\gamma } + p_B - p_N  - p_d \big)  \langle s_f | M ^{rc} | s_i \rangle    \nonumber \\
& \times \frac{1}{(2 \pi \hbar )^{6} }  \frac{1}{ \sqrt { 2 k _{\gamma 0}c ~2 E_B( \vec{p} _B )   2 E_N ( \vec{p} _N) 2 E_d( \vec{p} _d) } }  \hspace{0.0cm}, \\
 \langle f \vec{P} _f | S^{rc} | i \vec{P} _i \rangle &=   (-2 \pi i) \delta \big( k _{\gamma 0}c + E_B( \vec{p} _B ) - E_N ( \vec{p} _N) - E_d( \vec{p} _d) \big)   \nonumber \\
 &  \times  \langle f \vec{P} _f |  H^{e.m.}_I i0 G(E_i + i0) | i \vec{P} _i\rangle\hspace{0.0cm} ,
\\
 \langle f \vec{P} _f | S^{rc} | i \vec{P} _i \rangle & =  (-2 \pi i) \delta \big( k _{\gamma 0}c + E_B( \vec{p} _B ) - E_N ( \vec{p} _N) - E_d( \vec{p} _d) \big)   \nonumber \\
 &  \times \delta \big( \vec{k} _{\gamma} +  \vec{p} _B  -  \vec{p} _N -  \vec{p} _d \big) \frac{1}{c} \frac{ \sqrt{ 4 \pi  } \hbar c  }{ \sqrt{2 k _{\gamma 0 } c } } \frac{1}{(2 \pi \hbar ) ^{3/2}} \epsilon ^* _{\mu} ( \vec{k}_{\gamma} \lambda ) \nonumber    \\
 &   \times  \langle B |  j^{\mu} ( - \vec{k} _{\gamma}  , \vec{p} _B + \vec{p} _N + \vec{p} _d )  | \psi ^{( +)} _ {\alpha} ( \vec{q} _i ) \nu _  {\alpha _i } (Nd)  \rangle   | _ { \vec{q} _i   = \frac{ \vec{p} _d -  2 {\vec p}  _N } {3} } \hspace{0.0cm}; 
\end{align}
\reseteqn
$  \langle s_f | M ^{rc} | s_i \rangle $ is the singularity-free matrix element for radiative capture, from which the differential cross section
\blockeqn
\begin{gather}
\begin{align}  
\label{dcs}
 d \sigma ^{ rc } & =  \Big | \langle s_f | M ^{rc} | s_i \rangle \Big | ^2 \frac{dLips (p _N + p _d , k _{\gamma} , p _B ) }{ 4 c ^2  \sqrt { (p_N \cdot p _d ) ^2 - m^2 _N m^2_d c^4 }  }  
\end{align}              
\end{gather}
 is obtained. Its dependence on the spin projections $m_{s_i}$ and $M_{I_i}$ of nucleon and deuteron in the initial channel, collectively described by $s_i$, and on the helicity $\lambda $ and on the spin projection ${\mathcal M}_{B_f}$ of photon and trinucleon bound state in the final channel, collectively described by $s_f$, are explicitly indicated. $  \langle s_f | M ^{rc} | s_i \rangle $ is Lorentz-invariant and can be calculated in any frame. It is defined by
\begin{gather}
\begin{align}
   \langle s_f | M ^{rc} | s_i \rangle   = & \frac{ \sqrt{4 \pi } }{c} (2 \pi \hbar )^{3/2} \sqrt{2 E_B( \vec{p} _B ) 2 E_N ( \vec{p} _N) 2 E_d( \vec{p} _d) }  \nonumber  \\
&   \times \langle B |  j^{\mu} ( - \vec{k} _{\gamma}  , \vec{p} _B + \vec{p} _N + \vec{p} _d ) \epsilon ^* _{\mu} ( \vec{k}_{\gamma} \lambda ) | \psi ^{ (+) } _ {\alpha} ( \vec{q} _i ) \nu _  {\alpha _i } (Nd)  \rangle  | _ { \vec{q} _i   = \frac{ \vec{p} _d -  2 \vec{p} _N } {3} }  
\label{matrix}
\end{align}
\end{gather}
\reseteqn
according to \Eqs (\ref{smatrix}).
It contains our model description of the hadronic and e.m. interactions which 
 will --- for example with respect to experimental energies --- not be fully precise. The model description uses nonrelativistic energies for all hadrons involved. Since the matrix element in (\ref{matrix}) can be calculated in any frame, we choose the c.m. frame, i.e., $ \vec{p} _d = -  \vec{p} _N, \vec{p} _B = -  \vec{k} _{\gamma } $, and the following computational strategy: 
\begin{itemize}
 \item The experimental deuteron energy in the lab frame determines the total energy of the system and the relative momentum $ {\vec{q} _i}$ in the initial channel. This step is done using relativistic kinematics. 
  \item All hadron energies involved in the matrix element (\ref{matrix}) are chosen nonrelativistically and consistent with their model values. 
 \item Taking the computed trinucleon binding energy for calculating the rest mass $m _B $ and the average nucleon mass for the nucleon mass $m_N $, i.e., $m_N c ^2 = 938.919$ MeV, the magnitude of the photon momentum $|\vec{k} _{\gamma}|$ is determined employing the nonrelativistic forms (\ref{kineen}) for the hadron energies, i.e., $| \vec{k} _{\gamma}c|+ m_B c^2 +  {\vec{k} _{\gamma} ^2}/{6 m _N} = m_N c^2 + m_d c^2 + {\vec{q} _i ^2}/{(4 m_N/3)}   $. Since the model rest mass $m _B $ is not the experimental one, neither for \hethree{} nor for ${}^3$H reactions, the photon momentum does not have the experimental value.
\end{itemize}
The other building blocks for the differential cross section (\ref{dcs}) are the Lorentz-invariant phase-space element $ dLips(p_N + p_d , k _{\gamma},p_B ) $, i.e.,
\begin{gather}
\begin{align}
 & dLips(p_N + p_d , k _{\gamma},p_B )  =  \nonumber \\
& (2 \pi \hbar ) ^4  \delta  (  k _{\gamma} +  p _B  -   p _N -   p _d  ) \frac{d^3 k _{\gamma}}{ ( 2 \pi \hbar ) ^3 2 k_{\gamma 0} c} \frac{d ^3 p _B } { ( 2 \pi \hbar ) ^3  2 E_B( \vec{p} _B ) } \hspace{0.0cm} ,
\end{align}
\end{gather}
and the incoming flux $ 4 c ^2 \sqrt { (p_N \cdot p _d ) ^2 - m^2 _N m^2_d c^4 } $ . In contrast to the matrix element $ \langle s_f | M ^{rc} | s_i \rangle $ which contains our model assumptions, the kinematic parts of the differential cross section, the Lorentz-invariant phase-space element and the incoming flux, have to be calculated with the true experimental and relativistic energies of nucleon, deuteron and trinucleon bound state. In the c.m. system the differential cross section takes the form\blockeqn \onelabel{dcsfin}
\begin{gather} 
\frac{d \sigma ^{ rc } _{c.m.}}{d^2  { \hat {\vec{k}} _{\gamma}} } \Bigg | _{s_i \rightarrow s_f}  = \frac{1}{64 \pi ^2 } \frac{1}{ (\hbar c ) ^2  } \Big | \langle s_f | M ^{rc} | s_i \rangle \Big | ^2 
 \frac{ |\vec{k} _{\gamma} | }{ | \vec{p} _N  |} \frac{1} { \big (  E_N( \vec{p} _N ) + E_d ( \vec{p} _N ) \big ) ^2}  \hspace{0.0cm} .
\end{gather}
That form of the cross section still describes the transition between specified spin states. The spin-averaged differential cross section is 
\begin{gather} 
\overline{\frac{d \sigma ^{ rc } _{c.m.}}{d^2  { \hat {\vec{k}} _{\gamma}} }} = \frac{1}{6}\sum_{M_{N_i} M_{d_i} } \sum_{\lambda  M_{B_f}} \frac{d \sigma ^{ rc } _{c.m.}}{d^2  { \hat {\vec{k}} _{\gamma}} } \Bigg | _{s_i \rightarrow s_f} \hspace{0.0cm}.
\end{gather}
\reseteqn
\subsubsection{Two-Body Photo Disintegration of the Trinucleon Bound State}
\label{sec:phodis}
The definitions of the $ S $-matrix $\langle f \vec{P} _f | S^{dis} | i \vec{P} _i \rangle$, of the singularity-free matrix element $\langle s_f | M ^{dis} | s_i \rangle $ and of the differential cross section are similar to those of radiative capture in \Eqs (\ref{smatrix})--(\ref{dcsfin}). However, the initial and final states are interchanged and $  \epsilon ^* _{\mu} ( \vec{k}_{\gamma} \lambda )  $ is replaced by $ \epsilon  _{\mu} ( \vec{k}_{\gamma} \lambda ) $. The singularity-free matrix element is defined by
\blockeqn
\begin{gather}
\begin{align}
 \label{matrixphob}
&  \langle s_f | M ^{dis} | s_i \rangle   =  \frac{ \sqrt{4 \pi } }{c} (2 \pi \hbar )^{3/2} \sqrt{2 E_B( \vec{p} _B ) 2 E_N ( \vec{p} _N) 2 E_d( \vec{p} _d) }    \nonumber  \\
&  \times \langle\psi ^{( -)} _ {\alpha} ( \vec{q} _f ) \nu _  {\alpha _f } (Nd)  |  j^{\mu} (  \vec{k} _{\gamma}  , \vec{p} _N + \vec{p} _d +\vec{p} _B ) \epsilon  _{\mu} ( \vec{k}_{\gamma} \lambda ) | B  \rangle   | _ { \vec{q} _f   = \frac{ \vec{p} _d -  2 \vec{p} _N } {3} },
\end{align}
\end{gather}
 from which the differential cross section
\begin{gather}
\begin{align} 
\label{dcsdis}
 d \sigma ^{ dis }  =   \langle s_f | M ^{dis} | s_i \rangle \Big | ^2 \frac{dLips ( k _{\gamma} +p_B, p_N , p_d  ) }{ 4 c^2 (p_B \cdot k_ {\gamma})  }  \hspace{0.0cm} .
\end{align}
\end{gather}
\reseteqn
 is obtained. Its dependence on the helicity $\lambda $ and on the spin projection ${\mathcal M}_{B_i}$ of photon and trinucleon bound state in the initial channel, collectively described by $s_i$, and on the spin projections $m_{s_f}$ and $M_{I_f}$ of nucleon and deuteron in the final channel, collectively described by $s_f$, are explicitly indicated. $  \langle s_f | M ^{dis} | s_i \rangle $ is Lorentz-invariant and can be calculated in any frame. We choose the c.m. frame and follow the computational strategy of Sect. \ref{sec:rad} for radiative capture:
\begin{itemize}
 \item The experimental photon energy in the lab frame determines the total energy of the system and therefore the c.m. momentum $ \vec{k} _{\gamma}$ in the initial channel. This step is done using relativistic kinematics. 
  \item All hadron energies involved in the matrix element (\ref{matrixphob}) are chosen nonrelativistically and consistent with their model values. 
 \item Taking the computed trinucleon binding energy for calculating the rest mass $m _B $ and the average nucleon mass for the nucleon mass $m_N $, i.e., $m_N c ^2 = 938.919$ MeV, the relative momentum $ {\vec{q} _f}$ of the final nucleon-deuteron system is determined employing the nonrelativistic forms (\ref{kineen}) for the hadron energies, i.e., $ m_N c^2 + m_d c^2 + {\vec{q} _f ^2}/{(4 m_N/3)} = | \vec{k} _{\gamma}c|+ m_B c^2 +  {\vec{k} _{\gamma} ^2}/{6 m _N} $. Since the model rest mass $m _B $ is not the experimental one, neither for \hethree{} nor for ${}^3$H reactions, that nucleon-deuteron energy does not have the experimental value.
\end{itemize}
The other building blocks for the differential cross section (\ref{dcsdis}) are the Lorentz-invariant phase-space element $ dLips(  k _{\gamma} +p_B, p_N , p_d  )$, i.e.,
\begin{gather}
\begin{align}
&  dLips(  k _{\gamma} +p_B, p_N , p_d  )  =  \nonumber \\
&   (2 \pi \hbar ) ^4  \delta  (  k _{\gamma} +  p _B  -   p _N -   p _d  ) \frac{d^3 p _{N} }{ ( 2 \pi \hbar ) ^3 2 E_N( \vec{p} _N) } \frac{d ^3 p _d } { ( 2 \pi \hbar ) ^3  2 E_d( \vec{p} _d ) }  \hspace{0.0cm}, 
\end{align}
\end{gather}
and the incoming flux $ 4 c^2 (p_B \cdot k_ {\gamma}) $. In contrast to the matrix element $ \langle s_f | M ^{dis} | s_i \rangle $ which contains our model assumptions, the kinematic parts of the differential cross section, the Lorentz-invariant phase-space element and the incoming flux, have to be calculated with the true experimental and relativistic energies of nucleon, deuteron and trinucleon bound state. In the c.m. system the differential cross section takes the form 
\blockeqn 
\begin{gather}
\frac{d \sigma ^{ dis } _{c.m.}}{d^2  { \hat {\vec{p}} _N} } \Bigg | _{s_i \rightarrow s_f} = \frac{1}{64 \pi ^2 } \frac{1}{ (\hbar c ) ^2  } \Big | \langle s_f | M ^{dis} | s_i \rangle \Big | ^2 
 \frac{ | \vec{p} _N | }{ |\vec{k} _{\gamma}  |} \frac{1} { \big (  k _{\gamma 0}c + E_B(  \vec{k} _{\gamma } )   \big ) ^2}  \hspace{0.0cm} .
\end{gather}
That form of the cross section still describes the transition between specified spin states. The spin-averaged differential cross section is 
\begin{gather} 
\overline{\frac{d \sigma ^{ dis } _{c.m.}}{d^2  { \hat {\vec{p}} _N} } } = 
 \frac{1}{4} \sum_{\lambda  M_{B_i}} \sum_{  M_{N_f} M_{d_f} } \frac{d \sigma ^{ dis } _{c.m.}}{d^2  { \hat {\vec{p}} _N} } \Bigg | _{s_i \rightarrow s_f} \hspace{0.0cm} .
\end{gather}
\reseteqn

The $S$-matrix of photo disintegration and radiative capture are related, due to time reversal, in corresponding kinematics situations $| \vec{k} _{\gamma}|c + E_B( \vec{p} _B ) = E_N ( \vec{p} _N) + E_d( \vec{p} _d) $, i.e., $| \langle f \vec{P} _f | S^{dis} | i \vec{P} _i \rangle | = |\langle f \vec{P} _f | S^{rc} | i \vec{P} _i \rangle |  $, where the initial and final states of the matrix elements for photo disintegration and radiative capture are to be interchanged. That $S$-matrix relation is derived in \ref{time-rev} As a consequence, the differential cross section of  photo disintegration and radiative capture are obtained from each other by
\begin{equation}
\label{detbal}
\overline{\frac{d \sigma ^{ dis } _{c.m.}}{d^2  { \hat {\vec{p}} _N} }} \Bigg /\overline{ \frac{d \sigma ^{ rc } _{c.m.}}{d^2  { \hat {\vec{k}} _{\gamma}} } }  =  \frac{6}{4} \frac{ | \vec{p} _N |^2 }{ |\vec{k} _{\gamma}  |^2} \hspace{0.0cm} .
\end{equation}

\subsubsection{Spin Observables}
\label{sec:spinobs}
Spin observables are defined as in \Ref \cite{nemoto:98b}. In this paper we consider polarization in the nucleon-deuteron system only. Thus, analyzing powers $A$ in radiative capture are
\begin{equation}
\label{spinobs}
A = \frac{Tr \Big(  M ^{rc}    S^a M ^{rc \dagger}  \Big ) }{Tr \Big(  M ^{rc}  M ^{rc \dagger} \Big ) } \hspace{0.0cm} .
\end{equation}
The spin operators $S^a$ refer to the nucleon, i.e., $ S^a =  \{ \sigma _i \} $, or to the deuteron, i.e., $ S^a = \{ S_i , S_{jk} \} $; we take over the definitions of \Ref \cite{nemoto:98b}. For the deuteron analyzing powers the equivalent spherical tensor notation $iT_{11}$ and $T_{2m}$ will often be used as in \Ref \cite{nemoto:98b}. Details of the calculation of spin observables are also given in \ref{time-rev}

\label{kineenergy}
\renewcommand\themysection{\thesection}

\section{Results}
The results are based on calculations derived from the coupled-channel two-baryon potential A2 of \Ref \cite{hajduk:83a}. The potential is an extension of the purely nucleonic Paris potential~\cite{lacombe:80a} extended in isospin-triplet partial waves for single \diso~excitation. The \diso~is considered to be a stable particle of spin and isospin 3/2 with a rest mass $m_{\Delta}c^2$ of 1232 MeV. The extension is almost phase-equivalent~\cite{sauer:92a,picklesimer:91a} to the underlying nucleonic potential. The potential is used in the separably expanded form of \Ref \cite{nemoto:98a}. Partial waves up to total pair angular momentum $I = 2$ are taken into account.

The AGS three-particle equations for the trinucleon bound state $| B  \rangle$ are solved as in \Ref \cite{nemoto:98a}. The resulting binding energy $m_Bc^2 - 3m_Nc^2 $ is -7.702 MeV for the coupled-channel potential A2 and -7.373 MeV for the Paris potential, the purely nucleonic reference potential. If the Coulomb interaction is taken into account, as proper for \hethree, the binding energies shift to -7.033 MeV and -6.716 MeV; the e.m. shifts are calculated perturbatively in \Ref \cite{baier:82a}, amounting to 669 keV for a coupled-channel potential with \diso~excitation and to 657 keV for the purely nucleonic reference potential. Nevertheless, we use the purely hadronic values -7.702 MeV and -7.373 MeV as model values for calculating the matrix elements (\ref{matrix}) and (\ref{matrixphob}) of the considered photo reactions, since the calculation leaves out e.m. effects in the hadronic wave functions. For \hethree{}, as needed in most described reactions, they differ slightly from the experimental value -7.718 MeV. We shall describe proton-deuteron radiative capture for the two deuteron lab energies 19.8 MeV and 95 MeV; according to our calculational strategy, the photon energies in the final states are 12.05 (11.72) MeV and 36.97 (36.65) MeV for the theoretical binding energy of \hethree{} with (without) \diso, the discrepancy with the values 12.07 MeV and 36.99 MeV arising from the experimental \hethree{} binding energy is quite small. The wave function of the bound state is expanded in terms of the $({\mathcal LS})$ coupled basis states defined in Appendix~\ref{basicsta}; the expansion is done according to Appendix~\ref{boundsta}; the infinite expansion is first truncated to include all states up to total pair angular momentum $I=6$ in $(Ij)$ coupling and then transformed to $({\mathcal LS})$ coupling. The convergence of the presented results with respect to that truncation is checked and found to be satisfactory for all reactions corresponding to deuteron lab energies between 1 and 100 MeV: Additional wave function components usually change observables by less than 0.1\% in the complete kinematic regime with the exception of sensitive spin observables with values close to zero: They may change within 1\%.

The AGS three-particle equations for the nucleon-deuteron scattering states $| \psi ^{ (\pm)} _ {\alpha} ( \vec{q} ) \nu _  {\alpha}  (Nd) \rangle $ are solved as in \Ref \cite{chmielewski:98a99b}; in contrast to \Ref \cite{nemoto:98b}, \Ref \cite{chmielewski:98a99b} uses real-axis integration for the solution. The wave functions are expanded in terms of the $(Ij)$ coupled states defined in Appendix~\ref{basicsta}; the terms of the infinite expansion according to Appendix~\ref{scatsta}, contributing to the matrix element $ \langle\psi ^{( -)} _ {\alpha} ( \vec{q} _f ) \nu _  {\alpha _f } (Nd)  |j^{\mu} (  \vec{k} _{\gamma}  , \vec{p} _N + \vec{p} _d + \vec{p}_B ) \epsilon  _{\mu} ( \vec{k}_{\gamma} \lambda ) | B  \rangle $ of photo disintegration, are limited with respect to total three-particle angular momentum $\mathcal J$ and with respect to total pair angular momentum $I$ by the angular momenta contained in the trinucleon bound-state wave function and in the considered current multipoles. We note that also for radiative capture that matrix element of photo disintegration is calculated and time reversal is employed according to \ref{time-rev}

The e.m. current is taken over from \Ref \cite{strueve:87a}. Certain coupling constants, especially those referring to the \diso, are not up-to-date anymore; nevertheless, we decided not to change them in the present paper in order to preserve a valuable historic continuity with our past calculations~\cite{strueve:87a,henning:92a,sauer:94a} for related observables. The technique for calculating multipole matrix elements is developed in \Refs \cite{oelsner:99b,oelsner:99a}; a special stability problem arising in the calculation is discussed in Appendix~\ref{instab}. Our calculation for photo processes with internal excitation energies of the three-nucleon system up to $50$ MeV includes electric and magnetic dipole and quadrupole contributions, abbreviated in an obvious way by $E1, E2, M1$ and $M2$. The magnetic multipoles are calculated according to \Eq (\ref{vecharb}) of Appendix~\ref{curop} from the one- and two-baryon parts of the current. The electric multipoles are calculated in the Siegert form according to \Eq (\ref{tsig}) of Appendix~\ref{curop}. The first term of \Eq (\ref{tsig}) is calculated from the one- and two-baryon parts of the current. The second term of \Eq (\ref{tsig}) is the physically much more important proper Siegert term; it is replaced by model energies and Coulomb multipoles; the contributing charge density has diagonal single-nucleon and single-$\Delta$ contributions only; the nucleon-$\Delta$ transition contribution as well as two-baryon contributions are of relativistic order and are therefore omitted in the charge-density operator when calculating Coulomb multipoles. 
\subsection{Calculational Tests}
\subsubsection{Test of the Employed E.M. Current}

\setlength{\mylen}{0.95\textwidth}
\begin{figure}[htb]
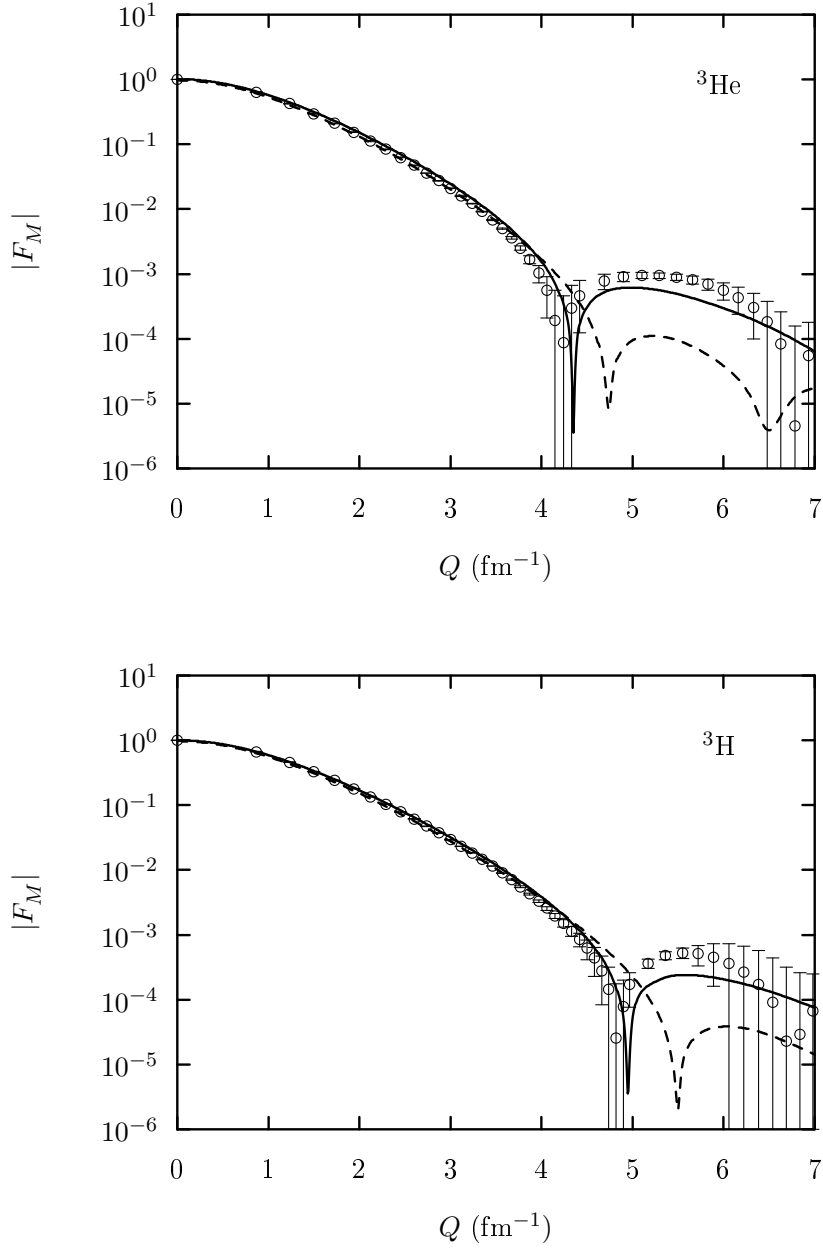

  \begin{center}
    \resizebox{0.9\hsize}{!}{\includegraphics{gmhek.10}}\\[10mm]
      \resizebox{0.9\hsize}{!}{\includegraphics{gmhhk.11}}
  \end{center}
  \caption{  Magnetic form factors $F_M$  of   ${}^3$He and ${}^3$H as
  function of momentum transfer $Q$. The results for the interaction with $\rmDelta $-isobar excitation
are shown as solid lines, whereas the results for the purely nucleonic
reference potential, the Paris potential, are shown as dashed lines. The experimental data are taken from \Ref \cite{amroun:94a} }\label{fig:gmh}
\end{figure}

\begin{figure}[htb]
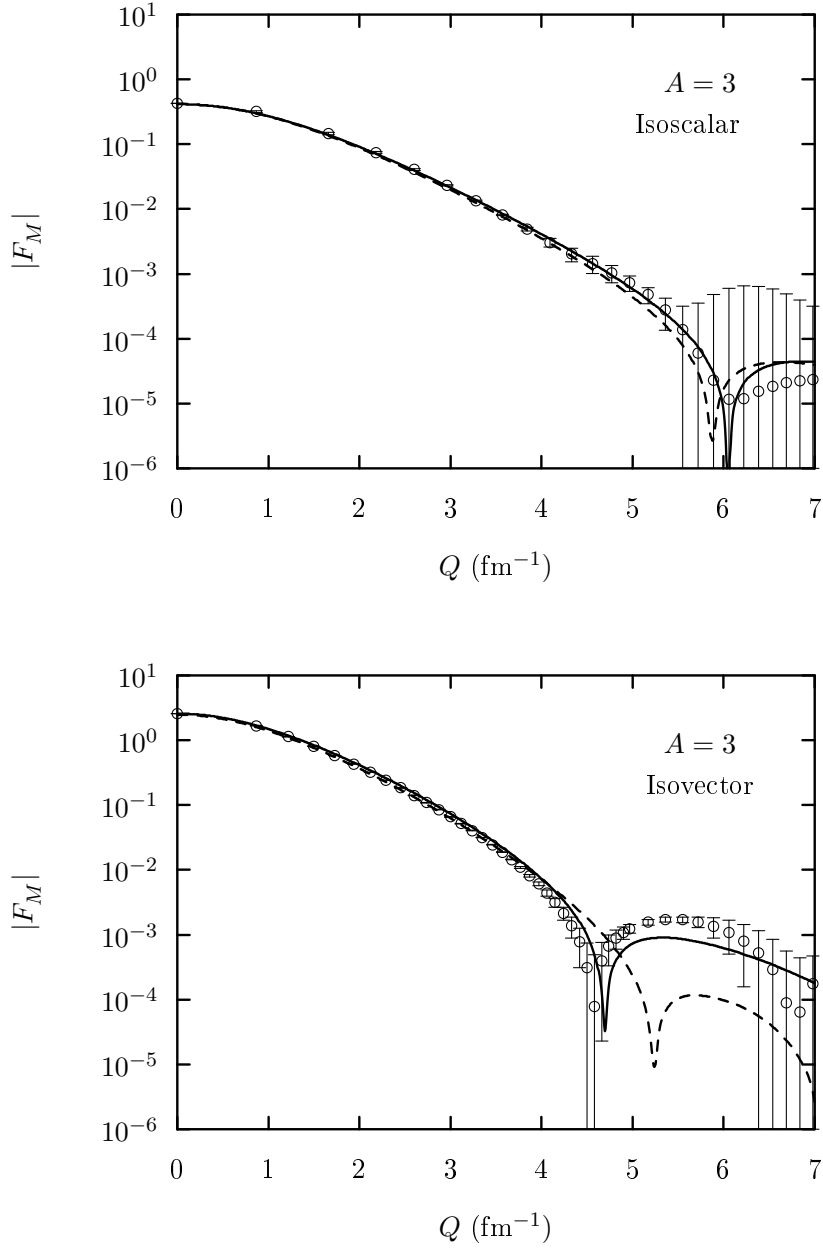

  \begin{center}
  \resizebox{0.9\hsize}{!}{\includegraphics{gmisk.12}}\\[10mm]
  \resizebox{0.9\hsize}{!}{\includegraphics{gmivk.13}}   
  \end{center}
  \caption{Isoscalar and isovector trinucleon magnetic form factors $F_M$ as
  function of momentum transfer $Q$. The results for the interaction with $\rmDelta $-isobar excitation
are shown as solid lines, whereas the results for the purely nucleonic
reference potential, the Paris potential, are shown as dashed lines. The experimental data are taken from \Ref \cite{amroun:94a} }\label{fig:gmsv}
\end{figure}
\setlength{\mylen}{0.75\textwidth}

As test of the employed e.m. current operator the magnetic form factors $F_M$ of the trinucleon bound state are calculated. In fact, the obtained results are recalculations of our previous ones given in \Refs \cite{strueve:87a,henning:92a,sauer:94a} with updated techniques; they are shown in \Figs \ref{fig:gmh} and \ref{fig:gmsv} individually for ${}^3$He and ${}^3$H and for their isospin-separated parts. Concerning the bound-state wave function, more mesh points and channels are included in the present calculation, significantly improving the results in and beyond the first diffraction minimum technically. Concerning the e.m. current, a sign mistake got detected in the two-baryon transition current from nucleonic to nucleon-$\Delta$ states; that contribution is in general very small, but becomes significant when all other contributions cancel, i.e., in the region of the diffraction minimum. Both effects combined, the one arising from the improved bound-state wave function and the other arising from a corrected two-baryon transition current, yield changes in the magnetic form factors at momentum transfer $Q > 4 \; {} \rm{fm}^{-1}$, decreasing the weight of the \diso~for the magnetic form factors somehow, compared with the results of \Ref \cite{sauer:94a}.
\setlength{\mylen}{0.75\textwidth}
\subsubsection{Test of the Calculational Apparatus for Radiative Capture}
\label{sec:teste1}
A detailed comparison of the calculational procedure of this paper and of the one in \Refs \cite{fonseca:93a,fonseca:00a} is carried out. The comparison is done for a purely nucleonic calculation and with a heavily truncated potential; only the two partial waves ${}^1S_0$ and ${}^3S_1-{}^3D_1$ are kept and the separable expansion is simplified to include three and four terms only, i.e., it is of rank 3 and 4, respectively. With respect to the current, only the $E1$ multipole is employed in the long-wave length form $T^{(1,l.w.)}_{elec\; m_1} (  k _{\gamma} / \hbar )$ according to \Eq (\ref{e1xlw}). Furthermore, the choice of momenta and energies for the current matrix element follows \Refs \cite{fonseca:93a,fonseca:00a} in the rest of this subsection and not our own computational strategy as discussed in Subsect.~\ref{sec:smatrix}. With the stated choice, the conceptional differences between the two calculations should be minimal and a detailed comparison is meaningful. On the other hand, the practical difference between the two calculations is enormous, especially with respect to the treatment of the permutation operator.

\begin{figure}[htb]
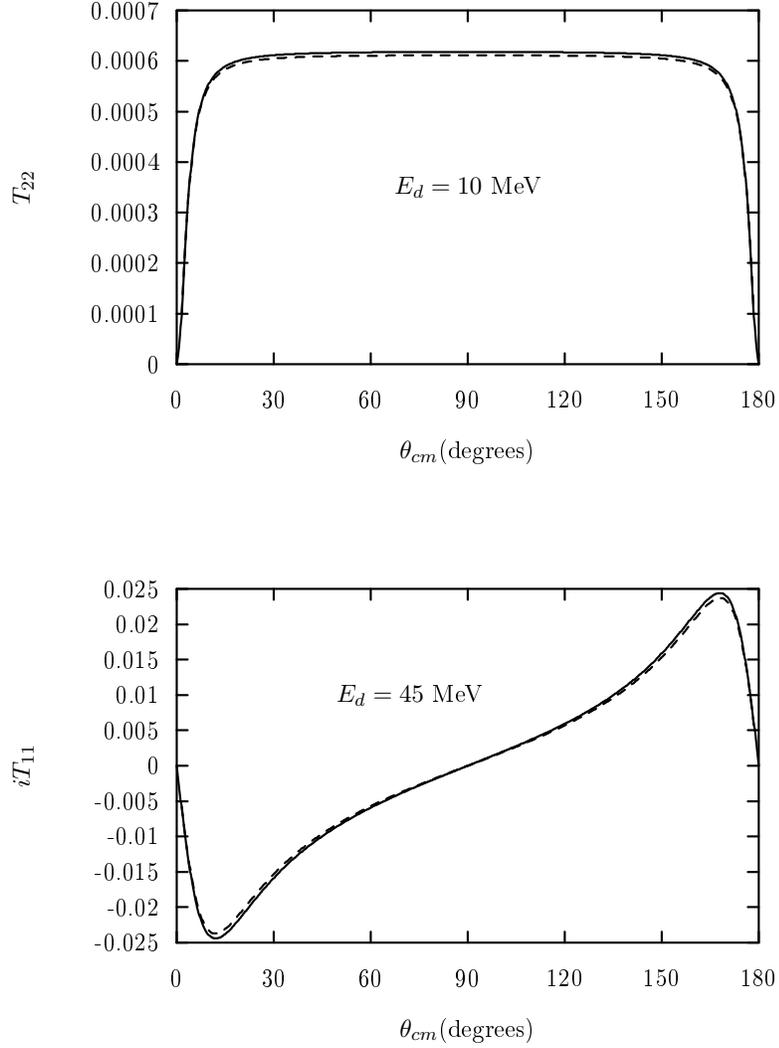

  \begin{center}

\resizebox{0.85\hsize}{!}{\includegraphics{t22.1}}\\[15mm]
\resizebox{0.85\hsize}{!}{\includegraphics{t1145.4}}
  \end{center}

  \caption{ \sloppy  Comparison of the computational procedure of this paper with the one of \Refs \cite{fonseca:93a,fonseca:00a}. The comparison uses the two spin observables $T_{22}$ and $iT_{11}$ of radiative capture at 10 MeV and 45 MeV  deuteron lab energies, respectively. The scattering angle $\theta_{cm}$ is the c.m. scattering angle of the photon with respect to the direction of the proton. The results use a simplified dynamics for the hadronic and e.m. interactions as described in Subsect.~\ref{sec:teste1}. The results based on the technique of this paper are shown as solid lines, the ones based on the technique of \Refs \cite{fonseca:93a,fonseca:00a} as dashed lines.} \label{fig:com1045k}
\end{figure}

The achieved agreement between the calculations of \Refs \cite{fonseca:93a,fonseca:00a} and of this paper is quite satisfactory. We have compared the differential cross section and spin observables $iT_{11},~{}T_{20},~{}T_{21},~{}T_{22}$ and $A_{yy}$. \Fig \ref{fig:com1045k} shows characteristic results. For 10 MeV we show the case of poorest agreement, for other observables the results of both calculations are not distinguishable in a plot. For 45 MeV we also show the poorest case of agreement; the differential cross section and $T_{22}$ have similar discrepancies, while the other observables are often not distinguishable in a plot. We conclude that our code for calculating trinucleon photo reactions is numerically quite reliable.

\subsubsection{Multipole Expansion of Current}
Our full calculations are based on a coupled-channel two-baryon potential with single \diso~excitation and on a corresponding e.m. current with the same coupling. The current is expanded into electric and magnetic multipoles. The convergence of results due to that multipole expansion of the current is studied in this subsection.
\begin{figure}[htb]
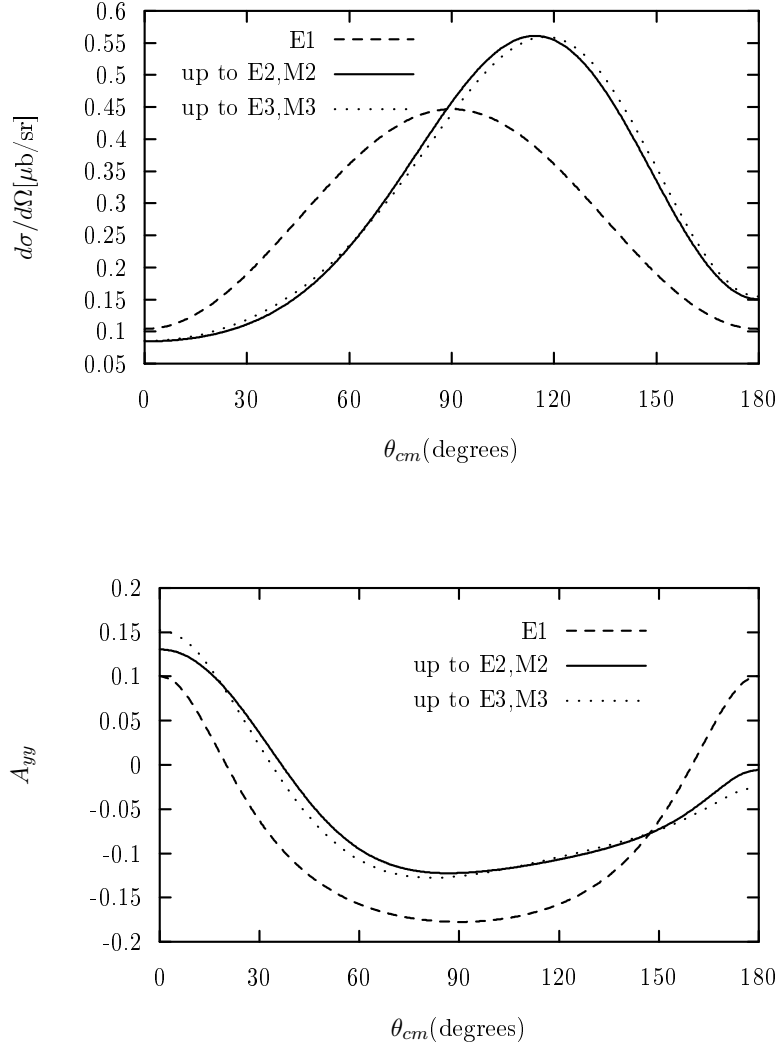

  \begin{center}
  \resizebox{0.85\hsize}{!}{\includegraphics{multconsec95.8}}\\[15mm]
   \resizebox{0.85\hsize}{!}{\includegraphics{multconayy95.5}}    
  \end{center}
  \caption{ Convergence of the multipole expansion of the e.m. current. The differential cross section and the spin observable $A_{yy}$
  are sample observables. Contributions from multipole up to $E3,M3$ are presented for radiative capture at 95 MeV deuteron lab
  energy as
  function of the c.m. scattering angle of the photon with respect to the direction of the deuteron; the $E1$ contributions include both terms of \Eq \eqref{tsig}. The results refer to full calculations with $\rmDelta $-isobar excitation.}\label{fig:multcon}
\end{figure}

The first study is carried out for the full calculations of radiative capture with $\rmDelta $-isobar excitation; the results for spin-averaged differential cross sections and for the spin observable $A_{yy}$ for 95 MeV, the highest considered deuteron lab energy, are shown in \Fig \ref{fig:multcon}. The electric dipole operator of the Siegert form, i.e., the second term in \Eq (\ref{tsig}) for $E1$, makes a significant contribution to most observables in the studied energy range. The first term of \Eq \eqref{tsig} for $E1$ is unimportant for the differential cross section, but sizeable for spin observables; this observed fact is not documented in \Fig \ref{fig:multcon}. The expansion with respect to the considered multipoles, i.e., multipoles up to $E2$ and $M2$, appears well converged. The contributions from the $E3$ and $M3$ multipoles, calculated tentatively, turn out to be very small for all observables, as is shown in \Fig \ref{fig:multcon}.
\begin{figure}[htb]
  \begin{center}
 \resizebox{0.85\hsize}{!}{\includegraphics{ayy90e1.0}}\\[15mm]
 \resizebox{0.85\hsize}{!}{\includegraphics{t2090e1.1}}
 \end{center}
\caption{Spin observables of radiative capture at $90^{\circ}$ lab scattering angle of the proton with respect to the direction of photon as function of deuteron lab energy. $A_{yy}$ is shown in the top, where the experimental data come from \Refs \cite{jourdan:85a} (open squares) and \cite{pitts:88a} (upper triangle); $T_{20}$ is shown in the bottom, where the experimental data come from \Refs \cite{schmid:96a} (open circles) and \cite{browne:96a} (cross). The results refer to calculations with the purely nucleonic reference potential. For the solid curves all multipoles up to $E2$, $M2$ are included, for the dashed curves only $E1$ in the Siegert form of term two in \Eq \eqref{tsig}, for the dotted curves its long-wavelength limit according to \Eq \eqref{e1xlw} and for the dashed-dotted curves $E1$ with both terms in \Eq \eqref{tsig}. In the top the solid and dotted curves are indistinguishable and are therefore not drawn separately.}\label{fig:mult-90deg}
\end{figure}

In \Fig \ref{fig:mult-90deg} the multipole expansion is studied for the energy dependence of the two spin observables $A_{yy}$ and $T_{20}$ at $90^{\circ}$ lab scattering angle of the proton. The study is carried out for calculations with the purely nucleonic
reference potential. Up to 30 MeV deuteron lab energy the electric dipole operator $E1$ in the Siegert form, i.e., the second term of \Eq \eqref{tsig}, determines the observables; in fact, its long-wavelength limit form \Eq \eqref{e1xlw} suffices. Beyond 30 MeV the first term in \Eq \eqref{tsig} for $E1$ and the $M1$ and $M2$ multipoles become significant, but they tend to cancel each other; this fact, which we consider a non-understood coincidence, makes the result based on the $E1$ operator in Siegert form close to the full result. The contributions of $E2$ and higher electric or magnetic multipoles remain small. In comparison with \Ref \cite{fonseca:00a} we note in passing that beyond 30 MeV deuteron lab energy also the improved hadronic interaction with respect to high partial waves and additional ranks in the separable expansion gives rise to rather sizeable changes.

\subsubsection{Frame Independence of Matrix Elements $ \langle s_f | M ^{rc} | s_i \rangle$ and $ \langle s_f | M ^{dis} | s_i \rangle$ in \Eqs \eqref{matrix} and \eqref{matrixphob} }
The current matrix elements  $ \langle s_f | M ^{rc} | s_i \rangle$ and $ \langle s_f | M ^{dis} | s_i \rangle$ are conceptionally Lorentz-invariant scalars and could therefore be calculated in any frame with identical results. We choose the c.m. frame for calculations of this paper. However, our calculational apparatus does not strictly respect Lorentz invariance, the results are frame-dependent. We therefore check that possible frame dependence of results and alternatively calculate the same current matrix elements also in the lab system. We do so for two-body photo disintegration at 36.9 MeV c.m. photon energy, i.e., for the matrix elements $ \langle s_f | M ^{dis} | s_i \rangle$. The recalculation in the 
lab system uses the initial lab photon momentum $\vec{k} _{\gamma}$ and requires the redetermination of relative momentum ${\vec{q} _f}$ of the final nucleon-deuteron system employing the nonrelativistic forms (\ref{kineen}) for the hadron energies, i.e., $ m_N c^2 + m_d c^2 + {\vec{q} _f ^2}/{(4 m_N/3)} +  {\vec{k} _{\gamma} ^2}/{6 m _N} = | \vec{k} _{\gamma}c|+ m_B c^2 $. The current operator needed for calculating the $S$-matrix element $ \langle s_f | M ^{dis} | s_i \rangle $ is of nonrelativistic order; it has the same functional form as for the calculation in the c.m. system, except for the single-baryon convection current.

The $E1$ contribution to the matrix element $ \langle s_f | M ^{dis} | s_i \rangle$ is calculated according to \Eq \eqref{e1xa}, the $E2$ contribution according to \Eq \eqref{vechara} and $M1$ and $M2$ contributions according to \Eq \eqref{vecharb}. Especially the two individual terms of the $E1$ contribution show a frame dependence of the order 2\%; however, that frame dependence cancels out for the sum of all multipoles considered. Resulting observables of two-body photo disintegration at 36.9 MeV c.m. energy are frame-independent up to a level of 0.01\%. Thus, we conclude that all matrix elements $ \langle s_f | M ^{rc} | s_i \rangle$ and $ \langle s_f | M ^{dis} | s_i \rangle$, calculated in this paper, are sufficiently frame-independent.
\clearpage
\subsection{Physics Results} 
\subsubsection{\Diso~Effects in Radiative Capture}
This subsection presents full results for radiative capture. They are compared with those of a purely nucleonic reference calculation without \diso~excitation which are derived from a purely nucleonic potential and a purely nucleonic current; the Paris potential is the nucleonic reference potential. 

\begin{figure}[b]
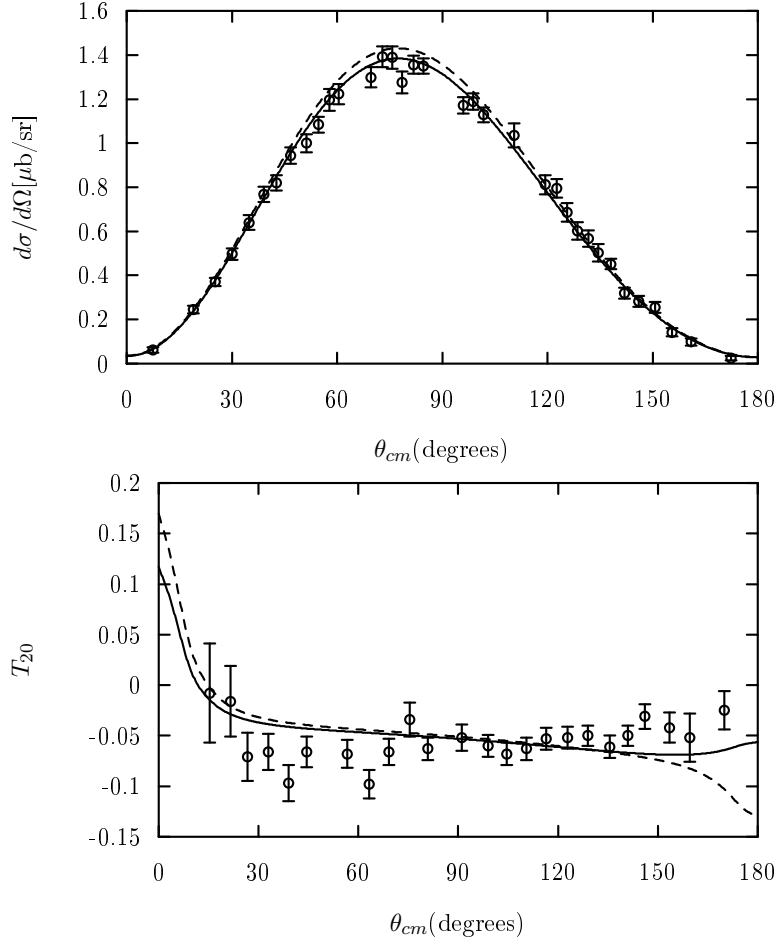

  \begin{center}
    \resizebox{0.85\hsize}{!}{\includegraphics{sec19.8.0}}\\[1mm]
    \resizebox{0.85\hsize}{!}{\includegraphics{t2019.8.3}}
  \end{center}

  \caption{ \sloppy  Radiative capture at 19.8 MeV deuteron lab
  energy as
  function of the c.m. scattering angle of the photon with respect to the direction of the proton. The differential cross section is shown on the top; the
  experimental  data are taken from \Ref \cite{belt:70a}. The spin observable $T_{20}$
  is shown on the bottom; the experimental data
  are taken from \Ref \cite{vetterli:85a}. The full results for the interaction with $\rmDelta $-isobar excitation
are shown as solid lines, whereas the results for the purely nucleonic
reference potential, the Paris potential, are shown as dashed lines.
 	} \label{fig:en19}
\end{figure}

\begin{figure}[htb]
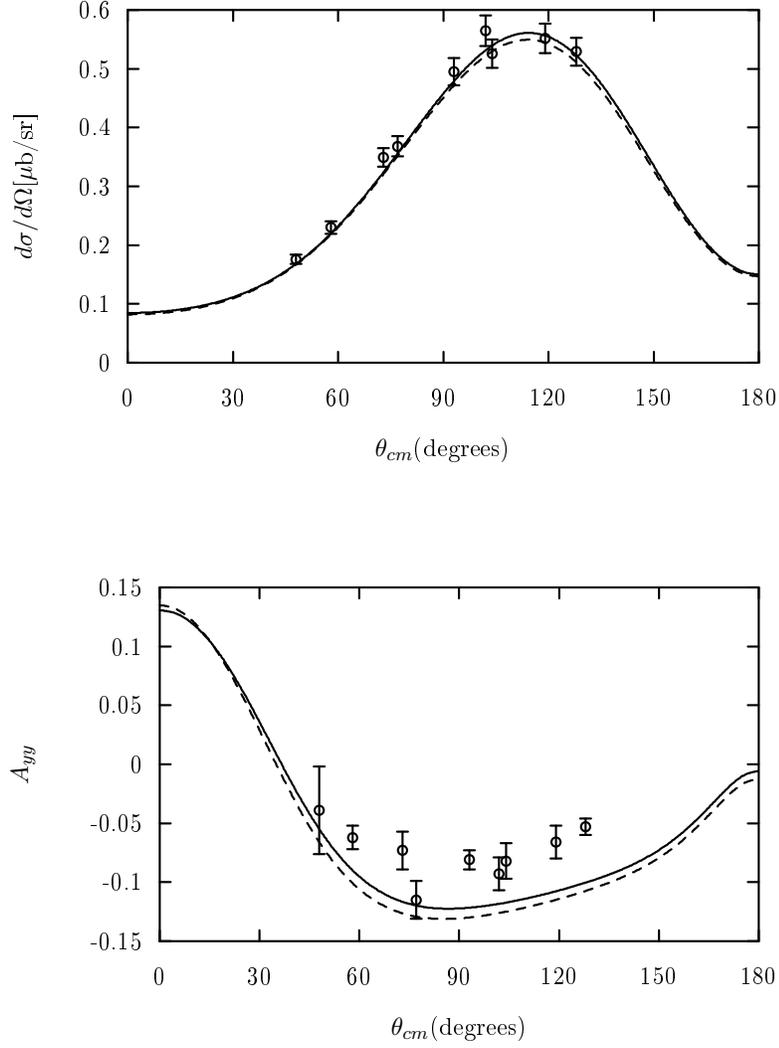

  \begin{center}
  \resizebox{0.85\hsize}{!}{\includegraphics{sec95.7}}\\[15mm]
   \resizebox{0.85\hsize}{!}{\includegraphics{ayy95.4}}
  \end{center}

  \caption{ \sloppy  Radiative capture at 95 MeV deuteron lab
  energy as
  function of the c.m. scattering angle of the photon with respect to the direction of the deuteron. The differential cross section is shown on the top. The spin observable $A_{yy}$
  is shown on the bottom; both sets of experimental data
  are taken from \Ref  \cite{pitts:88a}. The full results for the interaction with $\rmDelta $-isobar excitation
are shown as solid lines, whereas the results for the purely nucleonic
reference potential, the Paris potential, are shown as dashed lines.
 	} \label{fig:en95} 
\end{figure}

First, in \Figs \ref{fig:en19} and~\ref{fig:en95} we give sample results for 19.8 MeV and 95 MeV deuteron lab energies. In the figures we follow the standard convention for the presentation of data as introduced in \Refs \cite{belt:70a,vetterli:85a,pitts:88a}: At 19.8 MeV the c.m. photon angle refers to the direction of the proton, i.e., it is the c.m. \hethree{} angle with respect to the deuteron direction; at 95 MeV the c.m. photon angle refers to the direction of the deuteron.

%\vspace{-15cm}
%\vspace{15cm}

\begin{figure}[htb]
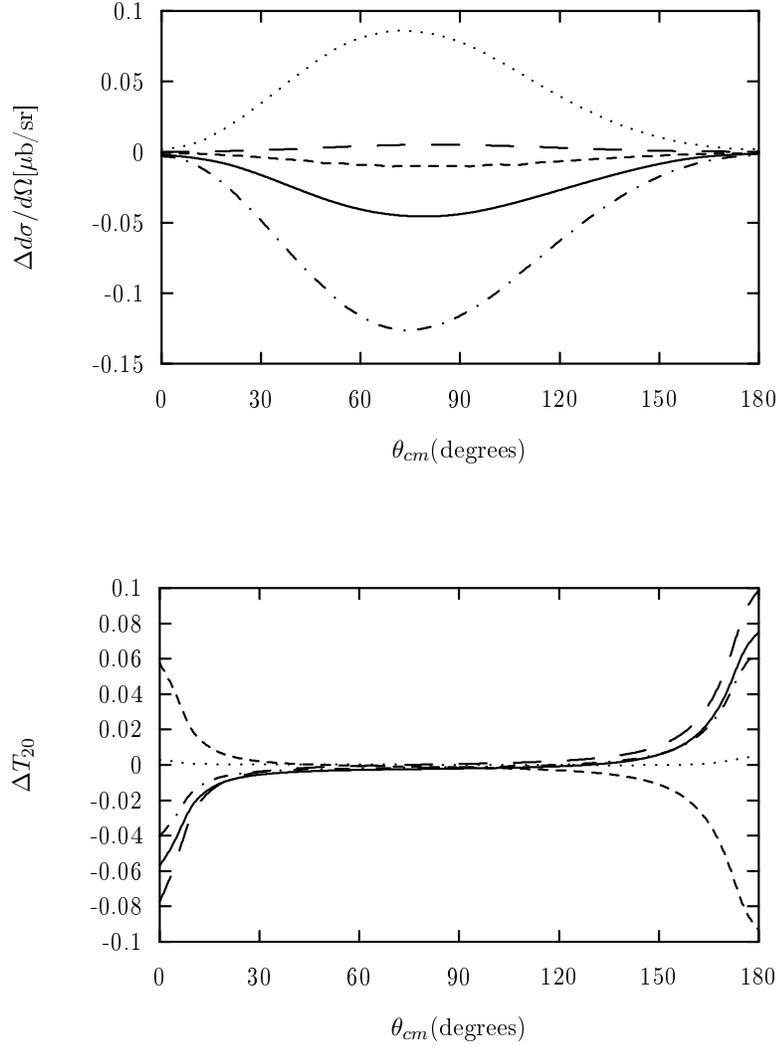

  \begin{center}
  \resizebox{0.85\hsize}{!}{\includegraphics{difdelsec19.8.1}}\\[15mm]
  \resizebox{0.85\hsize}{!}{\includegraphics{difdelt2019.8.3}}
  \end{center}

  \caption{ \sloppy   Contributions to the $\rmDelta $-isobar effect for radiative capture at 19.8 MeV deuteron lab
  energy as
  function of the c.m. scattering angle of the photon with respect to the direction of the proton. The c.m. differential cross section is shown on the top; the spin observable $T_{20}$
   on the bottom. Shifts in the observables are shown. The full shift in the prediction arising from the different predictions of the nucleonic reference potential and the coupled-channel potential is shown as solid lines. That full shift is split up into four parts, i.e., the contribution arising from the theoretical change of the triton binding energy due to scaling (dotted lines), the contribution arising from the change of the nucleonic components of the bound wave function (dashed-dotted lines), the contribution arising from the change of the nucleonic components of the scattering function (short-dashed lines), and the contribution arising from the  $\rmDelta $-components in both baryonic wave functions and the e.m. current (long-dashed lines); those four contributions add up to the full shift making up the complete  $\rmDelta $-isobar effect.	} \label{fig:delen19}
\end{figure}

\begin{figure}[htb]
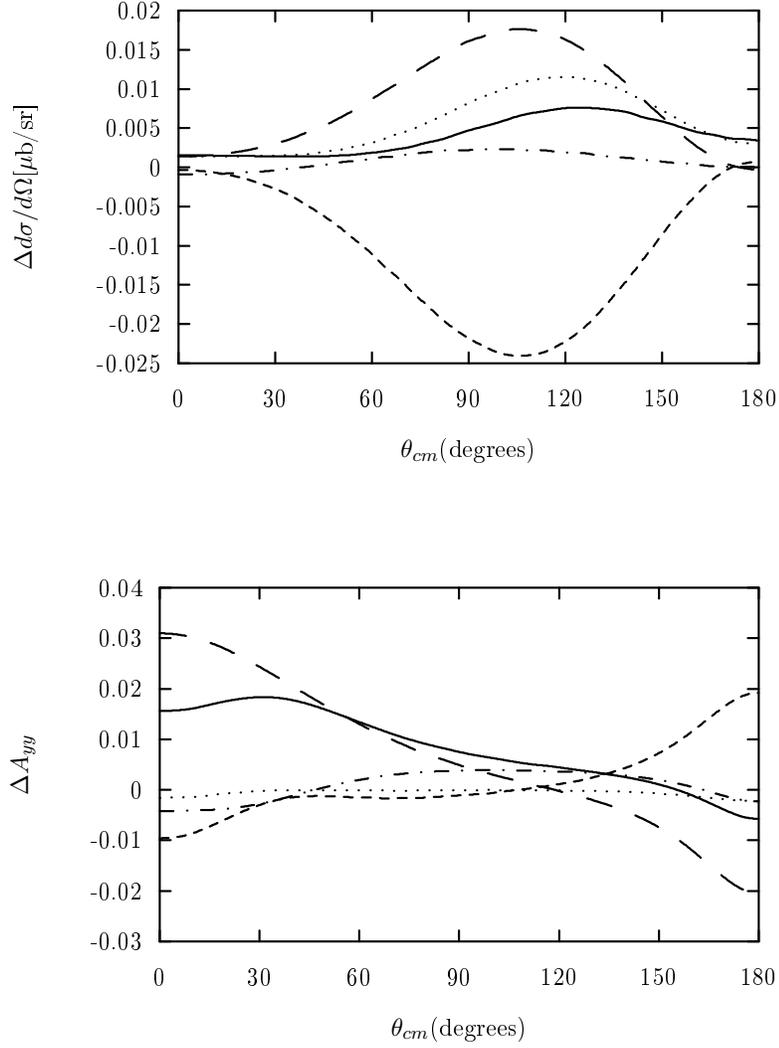

  \begin{center}
 \resizebox{0.85\hsize}{!}{\includegraphics{difdelsec95.9}}\\[15mm]
\resizebox{0.85\hsize}{!}{\includegraphics{difdelayy95.6}}
  \end{center}

  \caption{ \sloppy   Contributions to the  $\rmDelta $-isobar effect for radiative capture at 95 MeV deuteron lab
  energy as
  function of the c.m. scattering angle of the photon with respect to the direction of the deuteron. The differential cross section is shown on the top; the spin observable $A_{yy}$
  is shown on the bottom. Shifts in the observables are shown. The meaning of the different types of lines is the same as in \Fig \ref{fig:delen19}. } \label{fig:delen95} 
\end{figure}

The total \diso~effect is for all considered observables rather moderate. The inclusion of the \diso~does not change the phase-space or flux contributions to the cross sections, they are purely nucleonic and are given by the experimental conditions. However, the inclusion of the \diso~changes the current matrix element; its change can arise from four different sources:
\begin{itemize}
 \item The inclusion of the \diso~changes the theoretical trinucleon binding energy and thereby changes the theoretical photon energy of the process.
  \item The inclusion of the \diso~adds $\rmDelta$-components to the bound-state wave function and alters its nucleonic ones.
\item The inclusion of the \diso~adds $\rmDelta$-components to the wave function of the nucleon-deuteron scattering state and alters its nucleonic ones.
\item The inclusion of the \diso~adds current components which couple to $\rmDelta$-components in the hadronic wave functions.
\end{itemize}
\Figs \ref{fig:delen19} and~\ref{fig:delen95} try to distinguish the respective importance of those \diso~effects. The figures demonstrate that at the lower energy the \diso~effects come mainly from changes in the binding energy and wave function of the bound state, at the higher energy mainly from the scattering state and the $\rmDelta$-components of the e.m. current. 

Second, \Fig \ref{ob90deg} shows results for selected observables, i.e., $A_{yy}$ and $T_{20}$ at $90^{\circ}$ lab scattering angle of the proton, as function of deuteron lab energy, the \diso~effect is small. We remind the reader that in \Fig \ref{fig:mult-90deg} the multipole expansion is discussed for both observables. It is a physically remarkable effect, proven there, that below 30 MeV deuteron lab energy both observables are determined by the $E1$ Siegert operator of \Eq \eqref{e1xlw} of the long-wavelength limit in a model-independent way. In fact, the contribution of that simple operator describes the whole energy regime considered extremely well, but that result is above 30 MeV deuteron lab energy due to a cancellation of rather sizeable other contributions arising from the first term of the $E1$ operator in \Eq \eqref{tsig} and from other electric and magnetic multipoles. 

In conclusion, we have theoretical predictions for existing experimental data in the whole considered energy domain. However, we refrain from showing more results; they are available from the authors upon request.

\begin{figure}[htb]
  \begin{center}
 \resizebox{0.85\hsize}{!}{\includegraphics{ayytheta90new.0}}\\[15mm]
\resizebox{0.85\hsize}{!}{\includegraphics{t20theta90new.1}}
  \end{center}
  \caption{Spin observables of radiative capture at $90^{\circ}$ lab scattering angle of the proton with respect to the direction of photon as function of deuteron lab energy. $A_{yy}$ is shown in the top,where the experimental data come from \Refs \cite{jourdan:85a} (open squares) and \cite{pitts:88a} (upper triangle); $T_{20}$ is shown in the bottom, where the experimental data come from \Refs \cite{schmid:96a} (open circles) and \cite{browne:96a} (cross). The results for the full interaction with $\rmDelta $-isobar excitation
are shown as solid lines, whereas the results for the purely nucleonic
reference potential, the Paris potential, are shown as dashed lines.} \label{ob90deg}
\end{figure}
\clearpage

\subsubsection{Results for Two-Body Photo Disintegration}
According to \Eq (\ref{detbal}) the cross sections for photo disintegration and radiative capture are simply related. In fact, the experimental data as shown in the figures of this subsection are taken from radiative capture; 12.2 (36.9) MeV photon c.m. energy corresponds to 19.8 (95.0) MeV deuteron lab energy of radiative capture.

The relation of \Eq (\ref{detbal}) holds due to time reversal relation between experimental cross sections. Of course, time reversal relation remains true also for the matrix elements $M^{dis} $ and $M^{rc}$ of the model calculations, however, the meaning of kinematic correspondence is different in model calculations. We voted to always determine the initial c.m. momentum from the experimental lab energy by relativistic kinematics using experimental rest masses; the final c.m. momentum and energy are then determined by energy conservation using the nonrelativistic kinematics of the model calculation. Thus, the kinematics of the final state does not match properly the experimental situation. Our recipe, which is created by the fact that the model calculation is based on nonrelativistic quantum mechanics and is unable to reproduce the ${}^3$He binding precisely, clearly creates a conceptual uncertainty, the magnitude of which this subsection tries to estimate.

\begin{figure}[htb]
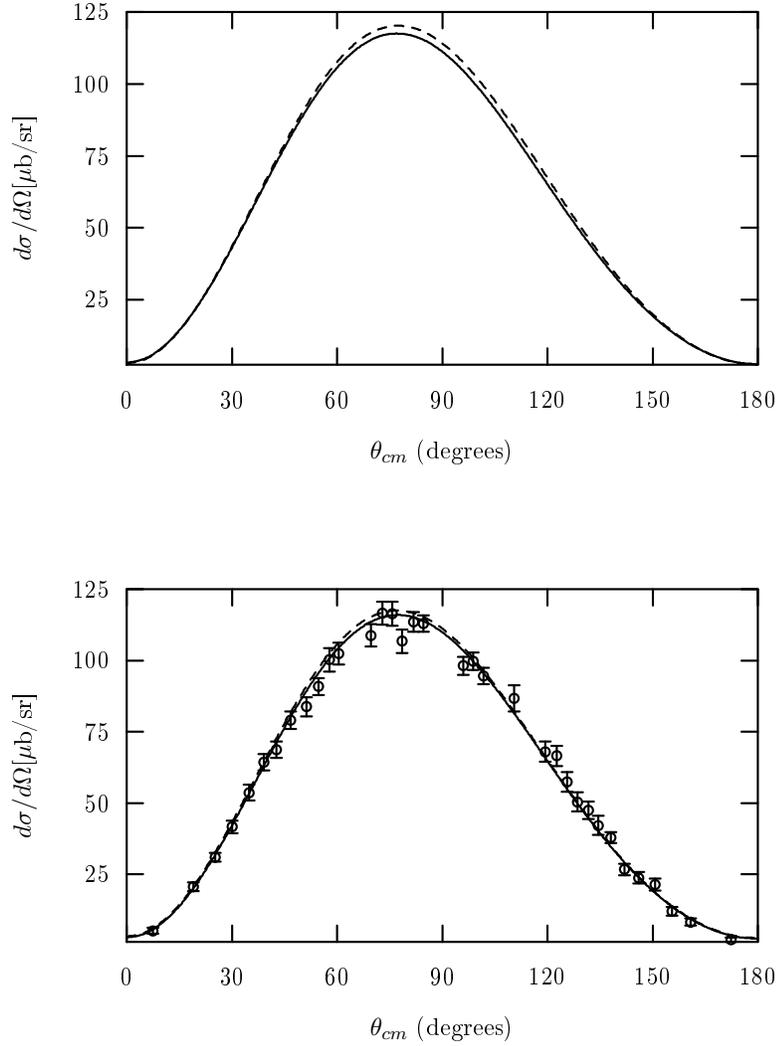

  \begin{center}
 \resizebox{0.85\hsize}{!}{\includegraphics{ecm12bal.0}}\\[15mm]
\resizebox{0.85\hsize}{!}{\includegraphics{ecm12.1}}     
  \end{center}

  \caption{ \sloppy  C.M. differential cross section of \hethree{} photo disintegration with two-body break-up at 12.2 MeV photon c.m. energy as
  function of the c.m. scattering angle of the proton with respect to the direction of the photon. (a) In the upper part, the results of computational methods based on detailed balance according to \Eq (\ref{detbal}) and on the  procedure of Subsect.~\ref{sec:phodis}, are compared by the dashed line and the solid line, respectively. The results refer to a purely nucleon reference calculation. (b) In the lower part of the figure, the complete \diso~effect is shown; here the calculations use the procedure of Subsect.~\ref{sec:phodis}. The full result for the coupled-channel interaction with \diso~excitation
is given by the solid line, the purely nucleonic reference result by the dashed line. } \label{fig:bal12} 
\end{figure}
\begin{figure}[htb]
  \begin{center}
  \resizebox{0.85\hsize}{!}{\includegraphics{ecm36bal.2}}\\[15mm]
  \resizebox{0.85\hsize}{!}{\includegraphics{ecm36.3}}  
  \end{center}

  \caption{ \sloppy  C.M. differential cross section of \hethree{} photo disintegration with two-body break-up at 36.9 MeV photon c.m. energy as
  function of the c.m. scattering angle of the deuteron with respect to the direction of the photon. (a) In the upper part, the results of computational methods based on detailed balance according to \Eq (\ref{detbal}) and on the  procedure of Subsect.~\ref{sec:phodis}, are compared by the dashed line and the solid line, respectively. The results refer to a purely nucleon reference calculation. (b) In the lower part of the figure, the complete \diso~effect is shown; here the calculations use the procedure of Subsect.~\ref{sec:phodis}. The full result for the coupled-channel interaction with \diso~excitation
is given by the solid line, the purely nucleonic reference result by the dashed line.} \label{fig:bal36} 
\end{figure}

We therefore calculated the spin-average differential cross section of \hethree{} two-body photo disintegration in two ways. In the first way, the experimentally corresponding radiative capture process is calculated and \Eq (\ref{detbal}) is employed for the step to two-body photo disintegration. In the second way, two-body photo disintegration is calculated directly according to the procedure of Subsect.~\ref{sec:phodis}. The results are compared in \Figs \ref{fig:bal12} and \ref{fig:bal36} for a purely nucleonic reference calculation, since the purely nucleonic reference calculation misses the \hethree{} binding most and therefore should show the largest differences between the two calculational procedures. The complete \diso~effect is also given in order to set the scale for the conceptual uncertainty of these calculations. Results are shown for 12.2 MeV and 36.9 MeV photon c.m. energies. Compared with the moderate complete \diso~effect the conceptual uncertainty is small at the larger photon energy, but sizable at the small one.

\begin{figure}[htb]
  \resizebox{0.85\hsize}{!}{\includegraphics{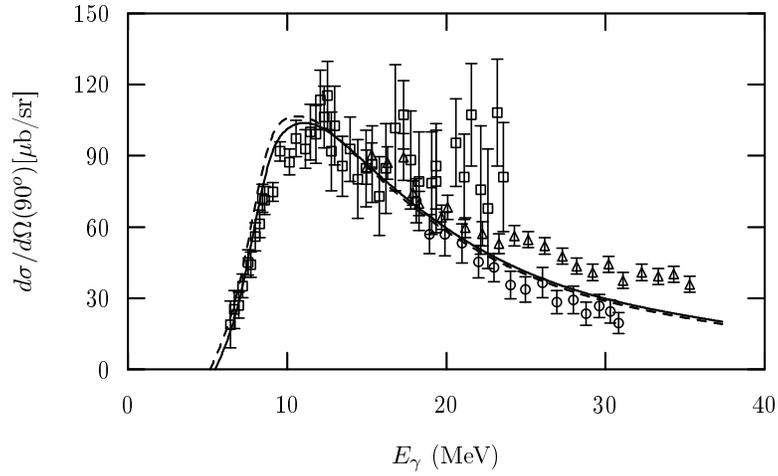}}
\caption{Lab differential cross section for {${}^3$H} photo disintegration with two-body break-up as function of the photon lab energy. The results refer to 90$^\circ$ nucleon lab angle with respect to the direction of the photon. The results of the full calculations  with $\rmDelta $-isobar excitation
are shown as solid lines, those of the reference calculations without $\rmDelta $-isobar excitation as dashed lines. Experimental data
  are taken from \Refs \cite{kosiek:66a} (open circles),\cite{faul:80a} (open squares), and \cite{skopik:81a} (upper triangles).} \label{tridis}
\end{figure}

In \Fig \ref{tridis} we show the energy dependence of the lab differential cross section for triton two-body photo disintegration with two-body break-up at 90$^\circ$ lab scattering angle of the nucleon with respect to the direction of the photon.

\section{Conclusion}
The paper calculates radiative nucleon-deuteron capture for selected energies and compares the theoretical predictions with existing proton-deuteron data. Some results for two-body photo disintegration are also given. The calculation includes the \diso~in the hadronic and the e.m. interactions. It employs fully correlated hadronic states and an e.m. current consisting of one-baryon and two-baryon contributions. 

The \diso~effect is isolated; it is moderate. At lower energies, it is mainly due to the shift in triton binding energy and due to the corresponding change in the nucleonic components of the bound-state wave function, brought about by the presence of the \diso. It is also found that below 30 MeV deuteron lab energy the spin observables $A_{yy}$ and $T_{20}$ at $90^{\circ}$ lab scattering angle can be obtained model-independently from the $E1$ Siegert term in the long-wavelength limit. At higher energies, the \diso~effect arises dominantly from the changes in the scattering wave function and in the e.m. current due to the \diso. 

%%\acknowledgments
\begin{acknowledge}
 L.~P. Yuan is supported in part by the DFG grant Sa 247/19-1 and by a graduate-student grant of Lower Saxony, M. Oelsner by the DFG grant Sa 247/20-1/2. J.~Adam Jr. is supported by the grant GA CR 202/00/1669. The authors are grateful to Th. Wilbois for important help with the technical apparatus at the beginning of this work. The calculations are performed at Regionales
Rechenzentrum f\"ur Niedersachen.
\end{acknowledge}

\begin{appendix}
\setcounter{equation}{0}
\renewcommand\themysection{\Alph{section}}
\section{Calculation of Amplitude}
\label{time-rev}
\begin{figure}[htb]
\hspace{-50pt}\psfig{file=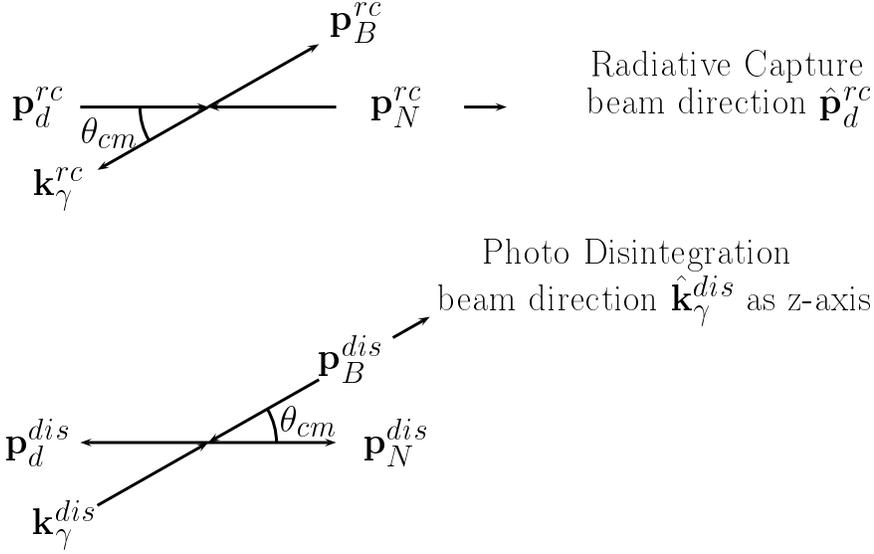,width=370pt,height=280pt}
\vspace{-50pt}
\caption{\sloppy Kinematic situations of nucleon-deuteron radiative capture and of two-body photo disintegration of the trinucleon bound state in the c.m. system. The relation between momenta connected by time reversal and the respective beam axes are shown.}
\label{fig:rev} 
\end{figure}

In Subsects. \ref{sec:rad} and \ref{sec:phodis} the calculation of cross sections for nucleon-deuteron radiative capture and of two-body photo disintegration of the three-nucleon bound state is described. Both reactions are related by time reversal ${\mathcal T}$. We use in Subsects. \ref{sec:rad} and \ref{sec:phodis} the same notation for the momenta of both reactions, though they are in fact different, i.e.,
\blockeqn
\begin{align}
{\mathcal T} | \vec{p}^{rc}_{N} \rangle = |(- \vec{p}^{rc}_{N}) \rangle = | \vec{p}^{dis}_N  \rangle \;\; , \\
{\mathcal T} | \vec{p}^{rc}_{d} \rangle = |(- \vec{p}^{rc}_{d}) \rangle = | \vec{p}^{dis}_d  \rangle \;\; , \\
 {\mathcal T} | \vec{k}^{rc}_{\gamma} \rangle = |(- \vec{k}^{rc}_{\gamma}) \rangle = | \vec{k}^{dis}_{\gamma}  \rangle  \;\; , \\
{\mathcal T} | \vec{p}^{rc}_{B} \rangle = |(- \vec{p}^{rc}_{B}) \rangle = | \vec{p}^{dis}_B  \rangle \;\; .
\end{align} 
\reseteqn
In this appendix we make the difference between the momenta of both reactions notationally explicit by the superscripts $rc$ and $dis$. Time-reversal also changes angular momentum eigenstates $ | jm \rangle $ by  ${\mathcal T} | jm \rangle =   (-1)^{j+m} |j(-m)\rangle$. Furthermore, the spin quantization axis, chosen in the experimental set-ups for the definition of spin observables, are different in both reactions. \Fig \ref{fig:rev} illustrates the kinematic situations of both reactions related by time reversal.

The current matrix element of radiative capture is obtained from the corresponding one of disintegration; the transverse helicities of the real photon are easier implemented in the disintegration matrix element choosing the direction of the incoming photon as z-direction.
\setlength{\bild}{0.895\linewidth}
\begin{align} 
\label{Ma}
  &      \langle B  {\mathcal M} _B |  j^{\mu} ( - \vec{k}^{rc} _{\gamma}  , \vec{p}^{rc} _B + \vec{p}^{rc} _N + \vec{p}^{rc} _d ) \overset{*}{\epsilon } _{\mu} ( \vec{k}^{rc}_{\gamma} \lambda ) | \psi ^{ (+) } _ {\alpha} ( \vec{q}^{rc}  ) \nu _  {\alpha  } M_{I}  m_{s} (Nd)  \rangle  \nonumber \\
\begin{split}
&  = (-)^{ \frac{1}{2}+{\mathcal M} _B } (-)^{ 1 + M_{I} } (-)^{\frac{1}{2} + m_{s}}  \\
&  \times \langle\psi ^{( -)} _ {\alpha} ( \vec{q}^{dis}  ) \nu _  {\alpha  } (-M_{I})  (-m_{s}) (Nd)  |  j^{\mu} (  \vec{k}^{dis} _{\gamma}  , \vec{p}^{dis} _N + \vec{p}^{dis} _d  + \vec{p}^{dis} _B  ) \epsilon  _{\mu} ( \vec{k}^{dis}_{\gamma} \lambda ) | B   ({-\mathcal M} _B) \rangle  \vspace{0.5em}.
\end{split}
\end{align}

The derivation of \Eq \eqref{Ma} uses the general properties of current and photon-field operators under time reversal, i.e.,
\blockeqn
\begin{align}
{\mathcal T}  J^{\mu}(x) {\mathcal T}^{-1} =  J_{\mu}({\mathcal T} x) \;\; , \\
{\mathcal T}  A^{\mu}(x) {\mathcal T}^{-1} =  A_{\mu}({\mathcal T} x) \;\; , 
\end{align}
and the condition
\begin{align}
 \overset{*}{\epsilon } {}^{\mu} (- \vec{k} \lambda ) = {\epsilon }_{\mu} (- \vec{k} \lambda ) \;\; .
\end{align} 
\reseteqn
 
In contrast to  Subsects. \ref{sec:rad} and \ref{sec:phodis} the projections of total deuteron and trinucleon angular momenta $ M_{I} $ and ${\mathcal M} _B$ as well as the projection of the nucleon spin $m_{s}$ are notationally indicated in the hadronic states. All spin projections in those states refer to the direction of the incoming photon momentum $\vec{k}^{dis} _{\gamma}$ in disintegration as quantization axis.

The only spin observables studied in this paper are analyzing powers in radiative capture. In contrast to the quantization axis used for the matrix elements \eqref{Ma}, the polarization of the deuteron and the nucleon are defined with respect to their respective momentum direction. Thus, either the spin dependence of the matrix element $ \langle s_f | M^{rc} | s_i \rangle $ has to be turned to the new quantization axis or the deuteron and nucleon spin operators $  \{ S_i , S_{jk} \} $ and $\sigma _i$ have to be transformed to them by the rotation $D(\alpha \beta \gamma ) = exp(-i \alpha J _z / \hbar ) exp(-i \beta J _y / \hbar ) \\ \times  exp(-i \gamma J _z / \hbar )$, $\vec J$ being the general angular momentum operator creating the rotation, i.e.,
\blockeqn
\begin{align}
& {S}^i  \longrightarrow  D(0 \theta _{cm} 0){S}^i D^\dagger (0 \theta _{cm} 0)  \;\; , \label{transd} \\
& {\sigma} _i \longrightarrow  D \big( 0 -(\pi - \theta _{cm}) 0 \big) {{\sigma} _i } D^\dagger \big( 0 -(\pi - \theta _{cm}) 0 \big)  \;\; .
\end{align} 
\reseteqn 
The transformation of $S_{jk}$ follows from \Eq \eqref{transd}.
With those transformations the spin observables \eqref {spinobs} of Subsect. \ref{sec:spinobs} are calculated. When the dependence of observables on the photon helicity and on the trinucleon spin are studied, no additional transformations of states or operators will be required.

\section{Details of Calculation}
\label{ap:details} 
This appendix calculates the current matrix element of two-body photo disintegration of the trinucleon bound state $  \langle\psi ^{ -} _ {\alpha} ( \vec{q} _f ) \nu _  {\alpha _f } (Nd)  |  j^{\mu} (  \vec{k} _{\gamma}  , \vec{p} _N + \vec{p}_d + \vec{p} _B ) \epsilon  _{\mu} ( \vec{k}_{\gamma} \lambda ) | B  \rangle $. The calculation is done in the c.m. system, i.e., $\vec{p} _d = -  \vec{p} _N, \vec{p} _B = - \vec{k} _{\gamma} $. The calculation uses a nonrelativistic potential model for the hadronic interaction. The baryon energies are employed in nonrelativistic forms.
\subsection{Basis States}
\label{basicsta}
The basis states  $ |pq \nu \rangle_{\alpha} $ describing the internal motion of three baryons are defined in Eq. (2.2) of \Ref \cite{nemoto:98a}. The discrete quantum numbers of the basis states, abbreviated by $ \nu $ are indicated in Fig. \ref{fig:jacobi}. The subscript $ \alpha $ at the states defines the role of particles as pair and spectator; this is why the subscript $ \alpha $ at quantum numbers is omitted, unless ambiguities could arise. In Fig. \ref{fig:jacobi} the $(Ij)$ coupling scheme is used, i.e.,
\begin{equation}
 \nu (I j)  = \{ [L(s_{\beta}s_{\gamma})S]I(l s)j \}  {\mathcal  J M_J}  
\{ (t_{\beta}t_{\gamma}) T t \}  {\mathcal   T M_T} Bb .
\end{equation}
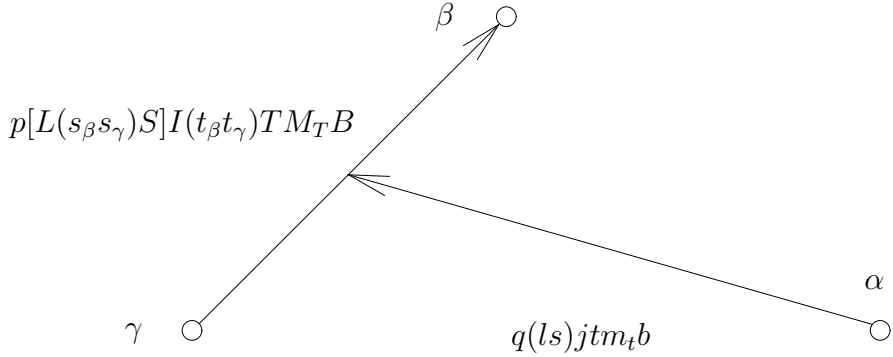
\begin{figure}[h]
  \centering \setlength{\unitlength}{0.00087489in}
\begingroup\makeatletter\ifx\SetFigFont\undefined%
\gdef\SetFigFont#1#2#3#4#5{%
  \reset@font\fontsize{#1}{#2pt}%
  \fontfamily{#3}\fontseries{#4}\fontshape{#5}%
  \selectfont}%
\fi\endgroup%
{\renewcommand{\dashlinestretch}{30}
\begin{picture}(6018,2104)(0,-10)
\put(1095,125){\ellipse{120}{120}}
\put(2985,2015){\ellipse{120}{120}}
\path(2039,1064)(5189,164)
\path(2286.249,1055.758)(2039.000,1064.000)(2253.282,940.375)
\path(2940,1970)(1140,170)
\path(2812.720,1757.867)(2940.000,1970.000)(2727.867,1842.720)
\put(5235,125){\ellipse{122}{122}}
\put(5142,372){\makebox(0,0)[lb]{\smash{{{\SetFigFont{12}{14.3}{\sfdefault}{\mddefault}{\updefault}$\alpha$}}}}}
\put(0,1317){\makebox(0,0)[lb]{\smash{{{\SetFigFont{12}{14.3}{\sfdefault}{\mddefault}{\updefault}$p [L(s_\beta s_\gamma)S]I (t_\beta t_\gamma) T M_T B$}}}}}
\put(3015,36){\makebox(0,0)[lb]{\smash{{{\SetFigFont{12}{14.3}{\sfdefault}{\mddefault}{\updefault}$q (ls)j t m_t b$}}}}}
\put(2558,1963){\makebox(0,0)[lb]{\smash{{{\SetFigFont{12}{14.3}{\sfdefault}{\mddefault}{\updefault}$\beta$}}}}}
\put(690,88){\makebox(0,0)[lb]{\smash{{{\SetFigFont{12}{14.3}{\sfdefault}{\mddefault}{\updefault}$\gamma$}}}}}
\end{picture}
}
  \caption{Three-baryon Jacobi momenta, and discrete quantum numbers. The spectator baryon is labeled $\alpha$, the pair is made up of baryons $\beta $ and $ \gamma $.}\label{fig:jacobi}
\end{figure}
The orbital angular momentum of the pair $L$ and of the spectator $l$ are first coupled with their respective spins $S$ and $s$ to total pair and spectator angular momenta $I$ and $j$, which are then combined to total angular momentum $ {\mathcal  J} $ with projection $ {\mathcal  M_J} $; the parity quantum number can be derived according to $\Pi = (-)^L (-)^l $. The isospin coupling of the pair isospin $T$ and of the spectator isospin $t$ is done correspondingly for total isospin $ {\mathcal  T} $ with projection ${\mathcal  M_T}$. In the trinucleon bound states and in nucleon-deuteron scattering without Coulomb interaction only basis states with total isospin ${\mathcal  T} = \frac{1}{2} $ are needed. The additional quantum numbers $(Bb)$ give the baryon characteristics of the pair and of the spectator, $B = 1(0)$ standing for a two-nucleon (nucleon-$\Delta $) pair, $ b=\frac{1}{2}(-\frac{1}{2} ) $   for a spectator nucleon ($\Delta $); baryon characteristics could be read off from the individual spin and isospin quantum numbers s and t; the additional quantum numbers $(Bb) $ are introduced for convenience. The discrete quantum numbers, distinct from the continuous Jacobi momenta $p$ and $q$, are abbreviated by $\nu (I j) $. 

Practical calculations suggest other coupling schemes for different parts of the numerics. The $( {\mathcal  L S } )$ coupling scheme
\begin{equation}
 \nu ({\mathcal   L S})  = \{ (Ll) {\mathcal  L} [(s_{\beta}s_{\gamma})S s] {\mathcal  S}  \}  {\mathcal  J M_J}  
\{ (t_{\beta}t_{\gamma}) T t \}  {\mathcal   T M_T} Bb
\end{equation}
and the channel-spin coupling scheme
\begin{equation}\label{eq:chan-spin}
 \nu  (lK )   =  \{ l ( [L(s_{\beta}s_{\gamma})S]Is ) K \}    {\mathcal  J M_J}  
\{ (t_{\beta}t_{\gamma}) T t \}  {\mathcal   T M_T} Bb 
\end{equation}
are also used. The transformation between coupling schemes 
\blockeqn  \onelabel{transcouple}
\begin{equation}
 \langle \nu ({\mathcal   L S} ) |  \nu (I j) \rangle = \hat{{\mathcal  L}} \hat{{\mathcal  S}} \hat{I} \hat{j} \left\{ \begin{array}{ccc}
 L & l & {\mathcal  L}   \\
 S & s & {\mathcal  S} \\
 I & j & {\mathcal J} \end{array}
  \right\}  ,
\end{equation}
\begin{equation}
 \langle \nu (lK ) |  \nu (I j) \rangle = (-)^{l+K+I+j} \hat{K} \hat{j} \left\{ \begin{array}{ccc}
 I & s & K \\
 l & {\mathcal J} & j \end{array}
  \right\}   , 
\end{equation}
\begin{gather}
\begin{split}
\langle \nu (lK) ({\mathcal   L S} ) |  \nu ({\mathcal   L S} ) \rangle = (-)^{L+S+s+K+{{\mathcal   S+J-L}} }\hat{{\mathcal  L}} \hat{{\mathcal  S}} \hat{I} \hat{K}  \\
  \times \left\{  \begin{array}{ccc}
 S & L & I \\
 K & s & {\mathcal  S}  \end{array}
  \right\}  \left\{ \begin{array}{ccc}
 L & {\mathcal  S} & K \\
 {\mathcal  J} & l & {\mathcal  L}  \end{array}
  \right\}  \hspace{0.0cm}  ,
\end{split} 
\end{gather} 
\reseteqn
using the abbreviation $\hat{a} = \sqrt{2a+1}$, are unitary. They do not involve isospin and baryon  content and are independent from the continuous Jacobi momenta $(pq)$ at which the transformations are carried out.

We calculate the matrix element $ \langle  \psi ^{ -} _ {\alpha} ( \vec{q} _f ) \nu _  {\alpha _f } (Nd)  |  j^{\mu} (  \vec{k} _{\gamma}  ,\vec{p} _B ) \epsilon  _{\mu} ( \vec{k}_{\gamma} \lambda ) | B  \rangle $ of photo disintegration first and then use time reversal for obtaining the corresponding one of radiative capture. The latter step is described in \ref{time-rev} The matrix element has three ingredients, i.e., the current operator \\$j^{\mu} (  \vec{k} _{\gamma}  ,\vec{p} _B ) \epsilon  _{\mu} ( \vec{k}_{\gamma} \lambda ) $, the initial bound state $|B  \rangle $ and the final scattering state  $ \langle  \psi ^{ -} _ {\alpha} ( \vec{q} _f ) \nu _  {\alpha _f } (Nd) | $. Since the three baryons can be considered identical, whereas the employed basis $ |pq \nu\rangle  _  {\alpha  } $ is not totally antisymmetric, explicit antisymmetrization has to be carried out with the permutation operator $P = P_{123} + P_{321} $. The permutation operator $P$ is described in \Ref \cite{nemoto:98b}.
\subsection{Three-Nucleon Bound State $ | B \rangle $ }
\label{boundsta}
The three-nucleon bound state $ | B \rangle $ is obtained from the corresponding Faddeev amplitude $  | \psi _ {\alpha} \rangle  $ according to
\blockeqn
\begin{equation} 
| B \rangle =   N(1+P) | \psi _ {\alpha} \rangle \hspace{0.0cm}
\end{equation}
$N $ being the normalization factor ensuring $ \langle B | B \rangle = 1$. The Faddeev amplitude $  | \psi _ {\alpha} \rangle  $ is best calculated in $(Ij)$ coupling, whereas for the computation of the current matrix element $ ({\mathcal  L} {\mathcal  S} ) $ coupling is the better choice, i.e.,
\begin{align}
 \langle pq \nu ({\mathcal  L} {\mathcal  S} ) |  B \rangle  = & N \int dp'dq'p'{}^2 q'{}^2 \sum _ { \nu ' ({\mathcal  L} '  {\mathcal  S} ' ) } \sum _ {I'j'}\langle pq  \nu ({\mathcal  L} {\mathcal  S} )  | (1+P) | p'q'  \nu ' ({\mathcal  L} '  {\mathcal  S} ' )   \rangle   \nonumber \\ 
  & \times \langle   \nu '({\mathcal  L} '  {\mathcal  S} '  )  |\nu ' ( I'j' ) \rangle \langle p' q' \nu ' ( I'j' ) | \psi _ {\alpha} \rangle .
\end{align} 
\reseteqn
\subsection{The Three-Nucleon Scattering State $| \psi ^{ (+)} _ {\alpha} ( \vec{q} ) \nu _  {\alpha} (Nd) \rangle $}
\label{scatsta}
The antisymmetrized nucleon-deuteron scattering state of internal motion \\ $| \psi ^{ (+)} _ {\alpha} ( \vec{q} ) \nu _  {\alpha} (Nd) \rangle $ is obtained from the corresponding plane-wave state \\ $ | \phi _ {\alpha} ( \vec{q} ) \nu _  {\alpha} (Nd) \rangle $ by the full resolvent $G(\lz)$, i.e.,
\blockeqn
\begin{align}
 | \psi ^{( +)} _ {\alpha} ( \vec{q} ) \nu _  {\alpha} (Nd) \rangle =&  i0G(E_{\alpha}(\vec{q} ) +i0 )(1+P)/\sqrt{3}  | \phi _ {\alpha} ( \vec{q} ) \nu _  {\alpha} (Nd) \rangle\hspace{0.0cm} , \\ 
| \psi ^{ (+)} _ {\alpha} ( \vec{q} ) \nu _  {\alpha} (Nd) \rangle =& (1+P)/\sqrt{3}\Big[ 1+G_0 (E+i0 ) T_{\alpha} (E+i0 ) G_0 (E+i0 ) U(E+i0 ) \Big]  \nonumber \\ 
 &  \times  | \phi _ {\alpha} ( \vec{q} ) \nu _  {\alpha} (Nd) \rangle |_{E = E_{\alpha}(\vec{q} )}\hspace{0.0cm} . \label{scatdefb}
\end{align}
\reseteqn
In \Eq \eqref{scatdefb} the two-baryon transition matrix $T_{\alpha}(Z)$ and three-particle multi-channel transition matrix $U(Z)$ are introduced; they are defined in \Ref \cite{nemoto:98a}. In this section, as in \Ref \cite{nemoto:98a}, the three-particle c.m. part of the kinetic energy operator is left out; $E_{\alpha}(\vec{q} ) $ is the internal nucleon-deuteron energy corresponding to the relative momentum $ \vec{q} $, i.e., $E_{\alpha} ({\vec q} ) = E_{Nd} ({\vec q} \vec{K} ) - {\vec{K} ^2}/({6 m _N})$ according to \Eq \eqref{enerkine}. The two-baryon interaction is separably expanded. Thus, the plane-wave state has the following form
\blockeqn
\begin{align} 
  \label{inita}
| \phi  _ {\alpha}&  ( \vec{q} )  \nu _  {\alpha} (Nd) \rangle  \nonumber \\
&=  G_0(E_{\alpha}(\vec{q} )+i0 ) |g^{(i_0 \pi _0 I_0 T_0) }_ {\alpha} M_{I} M_{T_0} \rangle  |   \vec{q}  s_0 m_{s} t_0 m_{t_0} b_0 \rangle _  {\alpha}  \hspace{0.0cm}  , 
    \\
  \label{initb}
 | \phi  _ {\alpha}& ( \vec{q} ) \nu _  {\alpha} (Nd) \rangle  \nonumber \\
 &= G_0 (E_{\alpha}(\vec{q} )+i0 ) \sum_{\Pi  {\mathcal  J} {\mathcal  M} _{{\mathcal  J}} 
  {\mathcal  T} {\mathcal  M} _{{\mathcal  T}} } \sum _{  jm_j lm_l  } | i_0 q \chi (I_0 j) \Pi {\mathcal  J} {\mathcal  M} _{{\mathcal  J}} 
  {\mathcal  T} {\mathcal  M} _{{\mathcal  T}} \rangle  _ {\alpha}  \nonumber \\
  & \times \langle  I_0 M_Ijm_j | {\mathcal  J} {\mathcal  M} _{{\mathcal  J}} \rangle   \langle  T_0M_{T_0}t_0m_{t_0} |{\mathcal  T} {\mathcal  M} _{{\mathcal  T}} \rangle  \langle lm_ls_0m_s|jm_j\rangle   Y^*_{lm_l}( \hat{\vec{q}} )
  \end{align}
\reseteqn
with the quantum numbers $ \pi _0 = +, I_0 = 1, T_0 = M_{T_0} = 0 $ and $s_0 = t_0 = b_0 = \frac{1}{2} $. The choice of the spin projections $M_I $ and  $m_s$ specifies experimental conditions, and the choice of the isospin projection $m_{t_0} = +  \frac{1}{2} $ makes the calculation refer to proton-deuteron scattering. Eq.(\ref{inita}) relates the deuteron state back to the corresponding form factor $  |g^{(i_0 \pi _0 I_0 T_0) }_ {\alpha} M_{I} M_{T_0} \rangle $ of the separable expansion; in Eq. (\ref{initb}) the partial-wave coupled states $| i_0 q \chi (I_0 j) \Pi {\mathcal  J} {\mathcal  M} _{{\mathcal  J}} 
  {\mathcal  T} {\mathcal  M} _{{\mathcal  T}} \rangle  _ {\alpha} $, introduced in Eq.(2.4) of Ref.\cite{nemoto:98b} for the description of nucleon-deuteron scattering, are used. The practical computation of the scattering state 
\blockeqn \onelabel{eqscatsta} 
\begin{align}
| \psi ^{ (+)} & _ {\alpha} ( \vec{q} ) \nu _  {\alpha} (Nd) \rangle = (1+P)/\sqrt{3}G_0(E+i0 )  \nonumber \\
\times & \Big[ 1+ |\vec{g} _ {\alpha} \rangle \vec{\ltau} _ {\alpha} (E+i0 ) \langle \vec{g} _ {\alpha} |  G_0 (E+i0 ) U(E+i0 ) G_0 (E+i0 ) \Big]\Big|_{E = E_{\alpha}(\vec{q} )}     \nonumber  \\
 \times & \sum_{\Pi  {\mathcal  J} {\mathcal  M} _{{\mathcal  J}} 
  {\mathcal  T} {\mathcal  M} _{{\mathcal  T}} } \sum _{  jm_j lm_l  } | i_0 q \chi (I_0 j) \Pi {\mathcal  J} {\mathcal  M} _{{\mathcal  J}} 
  {\mathcal  T} {\mathcal  M} _{{\mathcal  T}} \rangle  _ {\alpha}  \nonumber \\
  \times & \langle  I_0 M_Ijm_j | {\mathcal  J} {\mathcal  M} _{{\mathcal  J}} \rangle   \langle  T_0M_{T_0}t_0m_{t_0} |{\mathcal  T} {\mathcal  M} _{{\mathcal  T}} \rangle  \langle lm_ls_0m_s|jm_j\rangle   Y^*_{lm_l}( \hat{\vec{q}} )
%\label{eqscatsta}
\end{align}
\begin{align}\label{eqscatsta-b}
| \psi ^{ (+)}  _ {\alpha}& ( \vec{q}_{} ) \nu _  {\alpha} (Nd) \rangle = (1+P)/\sqrt{3}G_0(E+ i0 ) \nonumber \sum_{\Pi  {\mathcal  J} {\mathcal  M} _{{\mathcal  J}} 
  {\mathcal  T} {\mathcal  M} _{{\mathcal  T}} } \sum _{  jm_j lm_l  }    \nonumber \\
 \times & \Bigg[ | i_0 q_{} \chi (I_0 j) \Pi {\mathcal  J} {\mathcal  M} _{{\mathcal  J}} 
  {\mathcal  T} {\mathcal  M} _{{\mathcal  T}} \rangle  _ {\alpha} +  \nonumber \\
 & \sum_{\chi_c} \sum_{i'_c i_c} \int_0^\infty \!\! q_c^2dq_c    | i'_c q_c \chi_c(I_cj_c)\Pi {\mathcal  J} {\mathcal  M} _{{\mathcal  J}} 
  {\mathcal  T} {\mathcal  M} _{{\mathcal  T}} \rangle (i'_c| \ltau(E+ i0 , q_c \chi_c(I_cj_c)) | i_c) \nonumber   \\
    &  \hspace{2cm} \times  (i_c q_c \chi_c(I_cj_c) | X^{\Pi {\cal J T}}(E+ i0 ) | i q \chi(Ij) ) \Bigg] \Bigg|_{E = E_{\alpha}(\vec{q}_{} )}   \nonumber \\
\times & \langle  I_0 M_{I_{}}jm_j | {\mathcal  J} {\mathcal  M} _{{\mathcal  J}} \rangle   \langle  T_0M_{T_0}t_0m_{t_0} |{\mathcal  T} {\mathcal  M} _{{\mathcal  T}} \rangle  \langle lm_ls_0m_{s_{}}|jm_j\rangle   Y^*_{lm_l}( \hat{\vec{q}}_{} )
 \end{align} 
\reseteqn
 proceeds according to the strategy suggested by the separable expansion:
\begin{itemize}
 \item The operator $ G_0(E+i0)U(E+i0)G_0(E+i0) $ is computed between states $|iq \chi (Ij) \Pi {\mathcal  J} {\mathcal  M} _{{\mathcal  J}} 
  {\mathcal  T} {\mathcal  M} _{{\mathcal  T}} \rangle  _ {\alpha} $ in $(Ij)$ coupling; it and $\vec{\ltau} _ {\alpha} (Z)$ are diagonal in the three-particle quantum numbers $( \Pi {\mathcal  J} {\mathcal  M} _{{\mathcal  J}} 
  {\mathcal  T} {\mathcal  M} _{{\mathcal  T}} )$ and independent of $ {\mathcal  M} _{{\mathcal  J}} $ and  ${\mathcal  M} _{{\mathcal  T}}$; for $ G_0(E+i0)U(E+i0)G_0(E+i0)  $ half-off-shell elements are required. \Eq \eqref{eqscatsta-b} uses the matrix elements \\$(i_c q_c \chi_c(I_cj_c) | X^{\Pi {\cal J T}}(E+ i0 ) | i q \chi(Ij) )$ and $(i'_c| \ltau(E+ i0 , q_c \chi_c(I_cj_c)) | i_c)$  of $ G_0(E+i0)U(E+i0)G_0(E+i0) $ and of $ \vec{\ltau} _ {\alpha} (E+i0 )$ as introduced in \Ref \cite{nemoto:98b}.
\item The number of form factors $ | \vec{g} _ {\alpha} \rangle $, i.e., the assumption on the dynamics, truncates the number of arising partial waves $|iq \chi (Ij) \Pi {\mathcal  J} {\mathcal  M} _{{\mathcal  J}} 
  {\mathcal  T} {\mathcal  M} _{{\mathcal  T}} \rangle  _ {\alpha} $ for each total angular momentum $ {\mathcal  J} $. The truncation in $ {\mathcal  J} $ comes from the structure of the current, the matrix element of which is computed.
\item $ ( {\mathcal  LS} )$ coupled components $ {}_ {\alpha} \langle pq \nu ({\mathcal   L S} ) | \psi ^{( +)} _ {\alpha} ( \vec{q} ) \nu _  {\alpha} (Nd) \rangle $ of the scattering state are determined; those components are most convenient for the full calculation of the current matrix element. The permutation operator $P$ is applied in $(Ij)$ coupling and the result is then transformed to $( {\mathcal  LS} )$ coupling.
\end{itemize}

The three-nucleon scattering state $| \psi ^{( +)} _ {\alpha} ( \vec{q}  ) \nu _  {\alpha  } (Nd)  \rangle $ can be computed according to the steps explained in this subsection and the current matrix element $\langle\psi ^{( -)} _ {\alpha} ( \vec{q} _f ) \nu _  {\alpha _f } (Nd)  |  j^{\mu} (  \vec{k} _{\gamma}  , \vec{p} _N + \vec{p} _d +\vec{p} _B ) \epsilon  _{\mu} ( \vec{k}_{\gamma} \lambda ) | B  \rangle$ can be obtained from it and the bound state $ | B  \rangle$. We use that calculational avenue for numerical checks, but prefer a more direct alternative strategy for determing the current matrix element:
\blockeqn
\begin{align} \onelabel{eq:cur-n_ags}
& \langle \psi ^{ (-)} _ {\alpha} ( \vec{q}_{f} ) \nu _  {\alpha _f} (Nd) | j^{\mu} (  \vec{k} _{\gamma}  , \vec{p} _N + \vec{p} _d +\vec{p} _B ) \epsilon  _{\mu} ( \vec{k}_{\gamma} \lambda ) | B  \rangle \nonumber \\
& \hspace{20pt} = \langle \phi _ {\alpha} ( \vec{q}_{f} ) \nu _  {\alpha _f} |  \Omega (E + i0) (1+P)/\sqrt{3} j^{\mu} (  \vec{k} _{\gamma}  , \vec{p} _N + \vec{p} _d +\vec{p} _B ) \epsilon  _{\mu} ( \vec{k}_{\gamma} \lambda ) | B  \rangle  _{}\Bigg|_{E = E_{\alpha _f}(\vec{q}_{f} )} \label{eq:cur-n_agsa}\\
& \Omega (\lz)  =  1 + U(\lz)  G_0(\lz) T_{\alpha}(\lz) G_0(\lz),\label{eq:cur-n_agsb} \\
& \Omega (\lz)  =  \sum _{n = 0} ^ {\infty} \big( PT _ {\alpha}(Z)G_0(Z)\big)^n .\label{eq:cur-n_agsc}
\end{align} 
\reseteqn
\Eqs \eqref{eq:cur-n_ags} are consistent with \Eq \eqref{scatdefb} which relates the scattering state $| \psi ^{( +)} _ {\alpha} ( \vec{q} _i ) \nu _  {\alpha _i } (Nd)  \rangle $ to the multi-channel transition matrix $U\big(E_{\alpha _{}}(\vec{q}_{})+i0 \big) $. Since $U(Z)$ satisfies an integral equation \cite{nemoto:98a}, that property carries over to the current matrix element \eqref{eq:cur-n_agsa}, i.e.,
\begin{align} \label{newags-omega}
&\Omega (E+i0) (1+P)/\sqrt{3} j^{\mu} (  \vec{k} _{\gamma}  , \vec{p} _N + \vec{p} _d +\vec{p} _B ) \epsilon  _{\mu} ( \vec{k}_{\gamma} \lambda ) | B  \rangle  _{} \nonumber \\
& = (1+P)/\sqrt{3}j^{\mu} (  \vec{k} _{\gamma}  , \vec{p} _N + \vec{p} _d +\vec{p} _B ) \epsilon  _{\mu} ( \vec{k}_{\gamma} \lambda ) | B  \rangle  \nonumber \\
& \;\;\; + P T_{\alpha}(E+i0) G_0(E+i0) \Big[\Omega (E+i0) (1+P)/\sqrt{3}j^{\mu} (  \vec{k} _{\gamma}  , \vec{p} _N + \vec{p} _d +\vec{p} _B ) \epsilon  _{\mu} ( \vec{k}_{\gamma} \lambda ) | B  \rangle  _{}  \Big]
\end{align}
with $E =  E_{\alpha _f}(\vec{q}_{f} )$. The integral equation \eqref{newags-omega} has the same kernel $P T_{\alpha}(E+i0) G_0(E+i0)$ as the corresponding one for $U(E+i0)$ and can therefore be solved with the same nummerical techniques, just its driving term $(1+P)/\sqrt{3}j^{\mu} (  \vec{k} _{\gamma}  , \vec{p} _N + \vec{p} _d +\vec{p} _B ) \epsilon  _{\mu} ( \vec{k}_{\gamma} \lambda ) | B  \rangle $ is different. When taking the scalar product of the correlated state resulting from the integral equation \eqref{newags-omega} with the channel state $\langle \phi _ {\alpha} ( \vec{q}_{f} ) \nu _  {\alpha _f}(Nd) |$, the required current matrix element follows. This alternative calculational scheme, based on \eqref{newags-omega}, is the one usually employed for the numerics of the paper; it solves the integral equation for the scattering state $| \psi ^{( +)} _ {\alpha} ( \vec{q}  ) \nu _  {\alpha  } (Nd)  \rangle$ implicitly when forming the current matrix element.

\subsection{Current Operator} 
\label{curop}
We calculate the current matrix element of two-body disintegration of the trinucleon bound states for the current operator $j^{\mu} (  \vec{k} _{\gamma}  ,\vec{p} _N + \vec{p} _d + \vec{p} _B ) \epsilon  _{\mu} ( \vec{k}_{\gamma} \lambda ) $ in the c.m. system, i.e., $\vec{p} _N = - \vec{p} _d$, $\vec{p} _B = - \vec{k} _{\gamma}$, assuming that the momentum of the real photon defines the $z$ direction, i.e., $ \hat {\vec{k} } _{\gamma} = \hat{\vec e} _3$ and $ \lambda = \pm 1$. The current operator has the decomposition
\begin{align}
&j^{\mu} (   k _{\gamma}  \hat{\vec e} _3 ,  -{k} _{\gamma} \hat{\vec e} _3 ) \epsilon  _{\mu} ( {k}_{\gamma}\hat{\vec e} _3 \lambda ) \nonumber \\ 
&= - \frac{1}{\sqrt{2}} \sum _{j} \sqrt{4\pi} i^j \sqrt{2j+1} \Big[ T^{(j)}_{elec \lambda}(  k _{\gamma} / \hbar ) + \lambda  T^{(j)}_{magn \lambda}(  k _{\gamma} / \hbar )  \Big] 
 \label{curmult} 
\end{align} 
into electric and magnetic multipoles, i.e., $ T^{(j)}_{elec\;m_j}(  k  )$ and $ T^{(j)}_{magn\;  m_j}(  k  )$ with $k = k_{\gamma} / \hbar$ and $ k_{\gamma} =  |\vec{k}_{\gamma}| $ for the rest of this appendix. In the c.m. system the formal parametric dependence on the total momentum $\vec{K}' + \vec{K} = -\vec{k}_{\gamma}$ is dropped. The multipoles are defined by
\blockeqn \onelabel{multdef}
\begin{align}
\label{vechara}
  &T^{(j)}_{elec\; m_j}(  k  ) =  - \Big[\sqrt{\frac{j+1}{2j+1}} T^{([j-1,1]j)}_{ m_j} (k) + \sqrt{\frac{j}{2j+1}} T^{([j+1,1]j)} _{ m_j} (k) \Big] , \\
&T^{(j)}_{magn\; m_j}(  k  ) =  T^{([j,1]j)}_{ m_j} (k)
 \label{vecharb}
\end{align}
with
\begin{gather}
 \label{vechar}
 T^{([I,1]j)}_{ m_j} (k) =  \frac{(-i)^j}{4\pi} \int d^2 \hat{\vec{k}} \Big[ Y ^{(I)} (\hat{\vec{k}} ) \otimes j^{(1)} (\hbar \vec{k}) \Big]^{(j)}_{m_j}.
\end{gather}
\reseteqn
In Eq.(\ref{vechar}) $ Y ^{(I)} (\hat{\vec{k}} ) $ is the spherical tensor corresponding to the spherical harmonics $ Y_{Im_I}(\hat{\vec{k}}), j^{(1)} (\hbar \vec{k}) $ the spherical tensor corresponding to the spatial part of the corresponding current $ j^{\mu} (\hbar \vec{k},\vec{K}' + \vec{K}) $. The electric multipole operator is employed in the Siegert form
\blockeqn
\begin{align}
 &T^{(j)}_{elec\; m_j}(  k  ) =  - \sqrt{\frac{2j+1}{j}} T^{([j+1,1]j)}_{ m_j} (k) - \sqrt{\frac{j+1}{j}} \frac{1}{\hbar k} \Big[H,T^{(j)}_{coul\; m_j} (k) \Big]  \label{tsig}
\end{align}
with the Coulomb multipole
\begin{gather}
 T^{(j)}_{coul\; m_j} (k) =  T^{([j,0]j)}_{ m_j} (k)=  \frac{(-i)^j}{4\pi} \int d^2 \hat{\vec{k}}  Y _{jm_j} (\hat{\vec{k}} ) \rho (\hbar \vec{k}) \hspace{0.0cm} ,
\end{gather}
\reseteqn
$\rho (\hbar k) $ being the Fourier transform of the charge-density operator with normalization $\rho (0) = 1 $. The form (\ref{tsig}) follows from current conservation; the hadronic Hamiltonian $H$ includes the c.m. motion of the baryons. As is well known, the Siegert form of the electric multipole operator takes exchange two-baryon currents into account, even if the Coulomb multipole is derived from a charge density, composed of one-baryon operators only as done in the calculation of this paper.

For many low-energy photo processes the electric dipole makes the dominant contribution to the current.
\blockeqn
\begin{align} 
& \langle\psi ^{( -)} _ {\alpha} ( \vec{q} _f ) \nu _  {\alpha _f }(Nd)  |  T^{(1)}_{elec\; m_j} (  k _{\gamma} / \hbar ) | B  \rangle  \nonumber \\
& \;\; = \langle\psi ^{ -} _ {\alpha} ( \vec{q} _f ) \nu _  {\alpha _f }(Nd)  | (-)\sqrt{3}  T^{[2,1]}_{ m_j} (  k _{\gamma} / \hbar ) - \sqrt{2} c  T^{(1)}_{coul\; m_j} (  k _{\gamma} / \hbar )  | B  \rangle \label{e1xa} \\
& \;\; = \langle\psi ^{( -)} _ {\alpha} ( \vec{q} _f ) \nu _  {\alpha _f }(Nd)  |(-)\sqrt{3}  T^{[2,1]}_{ m_j} (  k _{\gamma} / \hbar )  \nonumber \\ 
& \hspace{35mm} -\sqrt{2} c  \int d^3 x j_{1} ( k _{\gamma}x / \hbar ) Y_{1m_j} (\hatbf{x} ) \rho (\vec{x} ) | B  \rangle   \hspace{0.0cm}. \label{e1xb}
\end{align} 
\reseteqn
That electric dipole operator is still exact. However, in applications it is often used in configuration space as in Eq.(\ref{e1xb}) and in the long-wavelength limit, $ k _{\gamma}x / \hbar \ll 1 $, in which the first term $ T^{[2,1]}_{ m_j} (  k _{\gamma} / \hbar ) $ can be neglect and the regular spherical Besel function of order 1, $j_{1} ( k _{\gamma}x / \hbar )$ is approximated by $ k _{\gamma}x /3\hbar $. Taking the charge operator $\rho (\vec{x} )$ as single-baryon operator
\blockeqn
\begin{align}
 \label{lwxop}
\rho (\vec{x} ) = \sum_i \frac{e_p}{2} \big( 1+\tau _3 (i) \big) \delta (\vec{x} - \vec{x} _i ) , 
\end{align}
the sum on $i$ extending over all baryons, $\tau _3 (i)$ being the isospin projection with values $\pm 1/2$ for nucleons and with values $\pm 3/2, \pm 1/2$ for a \diso, $e_p$ being the positive elementary charge, the long-wavelength electric dipole operator takes the form
\begin{align} 
& \langle\psi ^{ -} _ {\alpha} ( \vec{q} _f ) \nu _  {\alpha _f }(Nd)  |  T^{(1,l.w.)}_{elec\; m_1} (  k _{\gamma} / \hbar ) | B  \rangle  \nonumber \\
& \thickapprox    \frac{(-)}{\sqrt{4\pi}} \frac{ k _{\gamma}c}{\hbar} \langle\psi ^{ -} _ {\alpha} ( \vec{q} _f ) \nu _  {\alpha _f }(Nd)  | \sum_i \frac{e_p}{2} \big( 1+\tau _3 (i) \big) (x_i)^{(1)}_{m_1}  | B  \rangle \label{e1xlw}.
  \end{align}
Our calculation is in momentum space. Thus, the use of the configuration space operator (\ref{e1xlw}) is unnecessary. Nevertheless, we are able to simulate the use of the approximate operator (\ref{e1xlw}) by a trick:
\begin{itemize}
 \item Calculate the matrix element (\ref{e1xlw}) for a fictions unrealistically small photon momentum $ \mathcal{K}  _{\gamma} $ for which the approximation $j_{1} (  {\mathcal K} _{\gamma}x / \hbar )  \approx   {\mathcal K} _{\gamma}x /3\hbar $ with high precision.
\item Since the dependence on the physical photon momentum $ k _{\gamma} $ is linear in Eq.(\ref{e1xlw}) the proper matrix element for the operator (\ref{e1xlw}) is obtained by scaling with the factor $ k _{\gamma} / {\mathcal K} _{\gamma}  $, i.e.,
\begin{align} 
& \langle\psi ^{ -} _ {\alpha} ( \vec{q} _f ) \nu _  {\alpha _f }(Nd)    |  T^{(1,l.w.)}_{elec\; m_1} (  k _{\gamma} / \hbar ) | B  \rangle  \nonumber \\ 
&\; \;  =  \frac{ k _{\gamma} }{ {\mathcal K} _{\gamma} } \langle\psi ^{ -} _ {\alpha} ( \vec{q} _f ) \nu _  {\alpha _f }(Nd)    |  T^{(1)}_{elec\; m_1} (  {\mathcal K} _{\gamma} / \hbar ) | B  \rangle . 
\end{align}
\reseteqn

\end{itemize}
\subsection{Instability Problem in Calculation of Current Matrix Elements}
\label{instab}
The physics results of this paper depend on matrix elements of the e.m. current  $J^{\mu} ( \vec{Q})$ according to Eqs. (\ref{currentj}), (\ref{matrix}) and (\ref{matrixphob}). The computation of those current matrix elements is quite involved and will not be spelt out in detail in this appendix. The computation is based on the multipole decomposition of the current and on partial-wave decompositions of the initial and final baryonic states. The computation encounters particular numerical problems: The contributions arising from higher current multipoles and from higher angular momenta in the hadronic states get computationally split up into series of rather large numbers with alternating signs, the full results being comparatively small; the straight-forward evaluation of those series becomes instable. We are unable to cope with that instability properly and to avoid it for good. Nevertheless, we believe that the results, presented in this paper, are computationally quite reliable. This appendix tries to give convincing reasons for our claim.

The current has one- and two-body contributions. For computational purposes, each current  contribution is tensorially recoupled as tensor product of orbital, spin and isospin operators. The matrix elements of $ T^{([I,a]j)}_{ m_j}(k)$ with $k=k_{\gamma} / \hbar$ take the general form
\reseteqn
\begin{eqnarray}
\label{d11}
\lefteqn{
 \bra{\vec{p}\,'\vec{q} \,' }  
 T_{m_j}^{([Ia]j)}(k)
 \ket{ \vec{p} \vec{q}\; } } \nonumber\\
 &=&
 (-i)^j \;  (4\pi)^{\frac{1}{2}} \; 
 \sum_{\kappa} \; RK(a,b,e,j,I;\kappa) \;
 \int d^2 \hatbf{k}\;
  f_p (k, p_-, p_+, v_p) \; f_q (k, q_-, q_+, v_q) 
\nonumber\\
&&
\times \left( Y^{(\kappa)}(\hatbf{k}) \otimes 
         \left[ \fcal{O}^{(c)}(c_{p_+}, c_{p_-}, c_{p}, c_{q})  
                 \otimes S^{(d)} \right]^{(e)} 
 \right)^{(j)}_{m_j} , 
\end{eqnarray}
$\fcal{O}^{(c)}$ being the orbital and $S^{(d)}$ being the spin part. The matrix elements of  $T_{m_j}^{([Ia]j)}(k)$ are computed in partial-wave basis with the operators acting on the three-baryon bound state  $|B  \rangle $, i.e.,
\begin{eqnarray}
\label{wignereckart}
&& \hspace{-15mm} \bra{ {\mathcal J'M_{J'}} }
T_{m_j}^{([Ia]j)}(k)
\ket{B} \nonumber \\ 
\;\; &=&
(-)^{{\mathcal J'-M_J'}} 
\left(  \begin{array}{ccc}
         {\mathcal J'}           & j    & \frac{1}{2}  \\
        {\mathcal -M_{J'} }     & m_j  & {\mathcal M}_B 
        \end{array} \right) 
\langle {\mathcal J'} ||
T^{([Ia]j)}(k)
|| B \rangle,\phantom{aaaaa}
\end{eqnarray}
$1/2$ being the total angular momentum of the bound state and ${\mathcal M}_B$ its projection. For a two-body contribution the formal structure of the resulting matrix elements is as follows:
\begin{eqnarray}
 \label{d23}
\lefteqn{  \langle p' q' [(L'l')\fcal{L}' (S's')\fcal{S}'] {\mathcal J}' ||  
           T^{([Ia]j)}(k) || B \rangle }
\nonumber\\
&=&
 (-i)^j\; (4\pi)^{\thot}  \; 
 \sum_{\nu= \{ Ll \fcal{L} Ss \fcal{S} \}}\; 
 \langle (S's')\fcal{S}' || S^{(d)}  || (Ss) \fcal{S}  \rangle
 \phantom{ \left(  \begin{array}{c} 0  \\
                                    0   \end{array} \right) }    
 \nonumber\\ 
 &&
\times \sum_{\kappa} \; RK(a,b,e,j,I;\kappa) \;
 \sum_{\sigma} \; 
 RS(c,d,e,j;\fcal{L}',\fcal{S}',{\mathcal J}',\fcal{L},\fcal{S},\oot;\kappa,\sigma)
 \nonumber\\
 &&
 \times \sum_{\gamma_p \varphi_p \gamma_q \varphi_q} \; 
 (-)^{c + \sigma} \;
 \hat{\gamma_p}\; \hat{\varphi_p} \; \hat{\gamma_q}\; \hat{\varphi_q} \;
 \hat{c} \; \hat{\kappa} \; \hat{\sigma} \;
 \hat{\fcal{L}}' \; \hat{\fcal{L}} \; 
% 3j Symbol
\left(  \begin{array}{ccc}
         \varphi_p & \varphi_q & \kappa  \\
         0         & 0         & 0 
        \end{array} \right) 
\nonumber\\
&& 
% 9j Symbol
\times \left\{  \begin{array}{ccc}
         c_p         &  c_q       &  c      \\
         \varphi_p   & \varphi_q  & \kappa  \\
         \gamma_p    & \gamma_q   & \sigma 
        \end{array} \right\} \;
% 9j Symbol
\left\{  \begin{array}{ccc}
         L'         & L         &  \gamma_p \\
         l'         & l         &  \gamma_q \\
         \fcal{L}'  & \fcal{L}  &  \sigma   
        \end{array} \right\} \;
 \sum_{\rho} \;
 RK( c_{p}, c_{p_{-}}, c_{p_{+}}, \gamma_p, \varphi_p ; \rho) 
 \nonumber\\
 &&
\times \int\limits_{0}^{\infty} p^2 dp 
 \langle p' L' || P^{(\gamma_p)}(c_{p_+}, \rho, \varphi_p) || p L \rangle 
 \nonumber\\
 &&
\times \int\limits_{0}^{\infty} q^2 dq 
 \langle q' l' || Q^{(\gamma_q)}(c_q, \varphi_q, \varphi_q) || q l \rangle   
 \braket{pq\nu}{B}. 
\end{eqnarray}
Both formulae (\ref{d11}) and (\ref{d23}) are admittedly intransparent ones. The quantities $RK$ and $RS$ are of geometric nature and just combinations of coefficients like 3j and 6j symbols arising from several recouplings. However, the detailed forms of \eqref{d11} and \eqref{d23} are irrelevant for the following numerical consideration. 

In Eq. (\ref{d11}) the complete momentum dependence is buried in the two functions $ f_p (k, p_-, p_+, v_p) $ and $ f_q (k, q_-, q_+, v_q)$. The function $ f_p (k, p_-, p_+, v_p) $ is regular in $\vec k$, the limit $ k=0 $ can be carried out without any problem, making that function independent of $ \hatbf{k} $ and $ v_p  $. In contrast, the function  $ f_q (k, q_-, q_+, v_q)$ carries a $\delta$-function in  $\vec k$, which gets integrated out by acting the operator of Eq. (\ref{d11}) on the three-baryon bound state, i.e., by forming $\bra{\vec{p}\,'\vec{q} \,' }  
 T_{m_j}^{([Ia]j)}(k)
 \ket{ B} $; then the limit  $ k=0 $ can be performed leaving $ Y^{(\kappa)}(\hatbf{k})$ in  (\ref{d11}) and in  $\bra{\vec{p}\,'\vec{q} \,' }  
 T_{m_j}^{([Ia]j)}(k)
 \ket{ B} $ the only dependence on  $\hatbf k$. As a consequence the tensor order 
\begin{gather}
\hspace{0.5cm}\kappa = 
0, \hspace{5em} \text{if $k = 0$},
\label{kapeq0}
\end{gather}
results. In Eq. (\ref{d23}) the characteristics of the considered current contribution enters by the non-negative integer parameters $  c_{p}, c_{q} $ and $c$ with $ c_{q} = 0$ or $c_q = 1$. One numerically dangerous part of the expression is the
$q$-integration over the orbital operator $ Q^{(\gamma_q)}(c_q, \varphi_q, \varphi_q)$ which weighs the bound-state wave function $   
 \braket{pq\nu}{B}$. Close inspection, e.g., according to  Eq. (D.41) of \Ref \cite{oelsner:99a}, shows that the chosen computation carries powers of the ratios $q/k$ and $q'/k$ which for momentum transfer $k$ small compared with physically relevant $q$ and $q'$ in the baryonic wave functions become unwantedly large. The numerical instability arises from the integration of those orbital operators $ Q^{(\gamma_q)}(c_q, \varphi_q, \varphi_q)$ for which the contributing tensor order $\gamma_q $ becomes large. That source for instability is $p$-dependent and therefore contributes to the $p$-integration; it enhances instabilities in the integration over $ P^{(\gamma_p)}(c_{p_+}, \rho, \varphi_p)$, which arise independently. We note, however, from the internal structure of the $p$-integral on $ P^{(\gamma_p)}(c_{p_+}, \rho, \varphi_p)$ that in the limit $k=0$ the only contribution to $ P^{(\gamma_p)}(c_{p_+}, \rho, \varphi_p)$ is for vanishing $\varphi_p$, i.e.,
\begin{gather}
\hspace{0.5cm} \varphi_p = 0, \hspace{5em} \text{if $ k = 0$ } .
\label{phipeq0}
\end{gather}
The result follows from Eq. (D.14) of \Ref \cite{oelsner:99a}. Thus according to 
Eqs. (\ref{kapeq0}) and (\ref{phipeq0}) and according to the  $9j$-symbol in Eq. (\ref{d23}) $  \varphi_q = 0 $ as well as $ \gamma_q = c_q $ in the limit $ k = 0 $. Of course, outside the limit $k=0$ the tensor orders $ \kappa > 0, \varphi_p > 0, \varphi_q > 0 $ and $  \gamma_q \neq c_q  $ make contributions. Nevertheless, we observe a rapidly decreasing weight with increasing order  $\gamma_q $, i.e., we observe numerical convergence and stability well before instable fluctuations set in with high tensor orders $ \gamma_q $ of $ Q^{(\gamma_q)}(c_q, \varphi_q, \varphi_q)$. 

In the table we give one example for the procedure and the reached numerical stability. The sample observable is the isovector magnetic form factor $F_{M}(Q)$ of the three-nucleon bound state, the contribution arising from the $\pi$-contact current is given. The observable in the table corresponds to the one plotted in the lower part of \Fig \ref{fig:gmsv}. The momentum transfer is called $Q$ in that plot; it corresponds to the k in this appendix. The convergence with respect to the included maximal tensor order $\gamma_{q,max}$ of Eq. (\ref{d23}) is studied. Numerical instability sets in when the tensor order $ \gamma_{q,onset}$ is computed.

\begin{center}
\noindent
{\bf Table:} Contribution arising from the $\pi$-contact exchange current to the isovector trinucleon magnetic formfactor $F_{M}(Q) $
\vspace{0.5cm}

\footnotesize
\begin{tabular}{|l|ccccc|c|}
\hline
%\multicolumn{7}{|c|}{\text{ Isovector magnetic trinucleon formfactor $G_{M1}(k) $ from pi-contact current }}\\
%\hline
\multicolumn{1}{|c|}{ \raisebox{-7pt}[13pt][7pt]{$Q$ [fm$^{-1}$]} } &
\multicolumn{5}{|c|}{ \raisebox{0pt}[13pt][7pt]{$\gamma_{q,max}$ } } &
\multicolumn{1}{|c|}{ \raisebox{-7pt}[13pt][7pt]
{ $ \gamma_{q,onset} $} } 
\\
\raisebox{0pt}[13pt][7pt]{}       & 
\raisebox{0pt}[13pt][7pt]{$0$}         &
\raisebox{0pt}[13pt][7pt]{$1$}         &
\raisebox{0pt}[13pt][7pt]{$2$}         &
\raisebox{0pt}[13pt][7pt]{$3$}  &
\raisebox{0pt}[13pt][7pt]{$4$}  &
\multicolumn{1}{c|}{\raisebox{0pt}[13pt][7pt]{}}
\\
\hline
0.001 & -0.622761   &-0.678250  &-0.678250    &-0.678250   &  -0.678250   &  6 \\
0.01 & -0.622743   &-0.678231  &-0.678231    & -0.678231   &   -0.678231   & 8 \\
0.1 & -0.620939   &-0.676272  &-0.676289    &-0.676289    & -0.676289    & $>$ 8 \\
1.0 & -0.468388   &-0.510500  &-0.512039    & -0.511955   &  -0.511955   & $>$ 9 \\
\hline
\end{tabular}
\end{center}
\vspace{20pt}

\begin{figure}[h]
\hspace{-10pt}\psfig{file=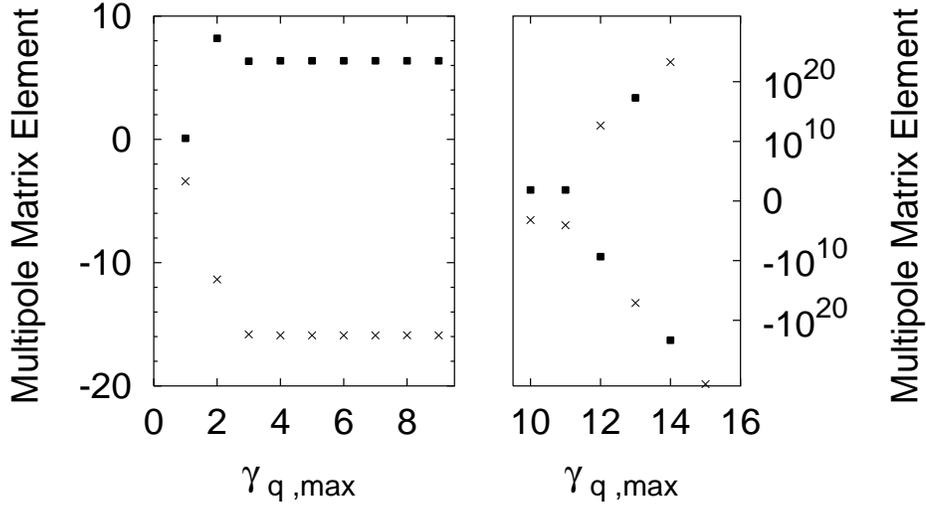,width=1.3\mylen}
\vspace{20pt}
\caption{Typical electric $E3$ (squares) and magnetic $M2$ (crosses) reduced multipole elements in units of fm${}^{3/2}$, as function  of  $\gamma_{q,max} $. The multipoles are derived from the two-nucleon $\pi $-contact current as in the table, i.e., they are isovectors with respect to their isospin character and they refer to radiative capture at 19.8 MeV deuteron lab energy. In the examples the matrix elements are calculated in the Wigner-Eckart reduced form with respect to total angular momentum and isospin for symmetrized plane-wave impulse approximation in the channel-spin coupling of \Eq \eqref{eq:chan-spin}, keeping the first term of \Eq \eqref{eqscatsta-b} only, i.e., without final-state interaction: $\langle   i_{0} q_{} \chi (l_f K_f) \Pi {\mathcal  J} 
  {\mathcal  T}   ||  G_0(E_f+i0) (1+P)/\sqrt{3}  T^{( j  )}_{elec (magn) }( {k} _{\gamma}/{\hbar})  ||B  \rangle$; the three-particle quantum numbers $( \Pi {\mathcal  J}  {\mathcal  T}) $ are chosen to be $\big((-)\;5/2\;1/2\big)$, the spectator quantum numbers $(l_fK_f) = (3\;1/2)$; for easier graphical presentation the multipole matrix elements are scaled by the numerical factor $1.5*10^{6}$.}\label{fig:conredmul}
\end{figure}

\Fig \ref{fig:conredmul} gives another example for the onset of instability in the calculation of typical multipole matrix elements needed for the description of photo processes. In the examples of the figure very high multipoles are chosen, since for them the instability problem becomes most severe. Though the examples are for special multipoles, special quantum numbers and for a final scattering state without final-state interaction, the on-set of instability is observed in the same general regime of $\gamma _q$ for all matrix elements and for the required realistic case in which final-state interaction in included. For lower multipoles the onset of instability occurs at somewhat higher $\gamma _q$. A corresponding on-set of instability is seen, when calculating current matrix elements for inelastic electron scattering. 

\end{appendix}

\bibliographystyle{fewbody}

\begin{thebibliography}{10}

\bibitem{nemoto:98a}
Nemoto, S., Chmielewski, K., Haidenbauer, J., Oryu, S., Sauer, P.~U.,
  Schellingerhout, N.~W.: Few-Body Systems {\bf 24},  213  (1998).

\bibitem{nemoto:98b}
Nemoto, S., Chmielewski, K., Haidenbauer, J., Meyer, U., Oryu, S., Sauer,
  P.~U.: Few-Body Systems {\bf 24},  241  (1998).

\bibitem{nemoto:98c}
Nemoto, S., Chmielewski, K., Oryu, S., Sauer, P.~U.: Phys. Rev. {\bf C58},
  2599  (1998).

\bibitem{chmielewski:98a99b}
Chmielewski, K., Fonseca, A., Nemoto, S., Sauer, P.~U.: Few-Body Systems
  Suppl. {\bf 10},  335  (1998).

\bibitem{gloeckle:96a}
Gl\"ockle, W., Wita{\l}a, H., H\"uber, D., Kamada, H., Golak, J.: Phys.\@
  Rep.\@ {\bf 274},  107  (1996).

\bibitem{witala:98a}
Wita{\l}a, H., Gl\"ockle, W., H\"uber, D., Golak, J., Kamada, H.:
  Phys.~Rev.~Lett. {\bf 81},  1183  (1998).

\bibitem{fonseca:93a}
Fonseca, A.~C., Lehman, D.~R.: Phys.~Rev. {\bf C48},  R503  (1993).

\bibitem{fonseca:00a}
Fonseca, A.~C., Lehman, D.~R.: Few-Body Systems {\bf 28},  189  (2000).

\bibitem{anklin:98a}
Anklin, H., de~Bever, L.~J., Buttazzoni, S., Gl\"ockle, W., Golak, J.,
  Honegger, A., Jourdan, J., Kamada, H., Kubon, G., Petijan, T., Qin, L.~M.,
  Sick, I., Steiner, P., Wita{\l}a, H., Zeier, M., Zhao, J., Zihlmann, B.:
  Nucl.~Phys. {\bf A636},  189  (1998).

\bibitem{golak:00a}
Golak, J., Kamada, H., Witala, H., Gloeckle, W., Kuros, J., Skibinski, R.,
  Kotlyar, V.~V., Sagara, K., Akiyoshi, H.: Phys. Rev. {\bf C62},  54005
  (2000).

\bibitem{yuan:00a}
Yuan, L.~P., Chmielewski, K., Oelsner, M., Sauer, P.~U., Fonseca, A.~C.,
  Adam~Jr., J.: Nucl.~Phys. {\bf A689},  433c  (2001).

\bibitem{strueve:87a}
Strueve, W., Hajduk, C., Sauer, P.~U., Theis, W.: Nucl.\@ Phys.\@ {\bf
  A465},  651  (1987).

\bibitem{adam:89a}
Adam~Jr.\@, J., Truhl\'{\i}k, E., Adamova, D.: Nucl.\@ Phys.\@ {\bf A492},
  556  (1989).

\bibitem{adam:91a}
Adam~Jr.\@, J., Hajduk, C., Henning, H., Sauer, P.~U., Truhl\'{\i}k, E.:
  Nucl.\@ Phys.\@ {\bf A531},  623  (1991).

\bibitem{henning:92a}
Henning, H., Sauer, P.~U., Theis, W.: Nucl.\@ Phys.\@ {\bf A357},  367
  (1992).

\bibitem{hajduk:83a}
Hajduk, C., Sauer, P.~U., Strueve, W.: Nucl.\@ Phys.\@ {\bf A405},  581
  (1983).

\bibitem{lacombe:80a}
Lacombe, M., Loiseau, B., Richard, J.~M., Vinh~Mau, R., Cot\'e, J., Pir\`es,
  P., de~Tourreil, R.: Phys.~Rev. {\bf C21},  861  (1980).

\bibitem{sauer:92a}
Sauer, P.~U.: Nucl.~Phys. {\bf A543},  291c  (1992).

\bibitem{picklesimer:91a}
Picklesimer, A., Rice, R.~A., Brandenburg, R.: Phys.\@ Rev.\@ {\bf C44},
  1359  (1991).

\bibitem{baier:82a}
Baier, H., Bentz, W., Hajduk, C., Sauer, P.~U.: Nucl.~Phys. {\bf A386},
  460  (1982).

\bibitem{sauer:94a}
Sauer, P.~U., Henning, H.: Few-Body Systems {\bf Suppl.\@ 7},  92  (1994).

\bibitem{oelsner:99b}
Oelsner, M., Adam~Jr.\@, J., \@, Sauer, P.~U., in: Proceedings of the 7th Conference Meson \& Light Nuclei, Prague-Pruhonice 1998 (Adam~Jr.\@, J., ed.), World Scientific 1999, p.486.

\bibitem{oelsner:99a}
Oelsner, M.: Ph.D. Thesis, University of Hannover 1999.

\bibitem{amroun:94a}
Amroun, A., \emph{et al.}: Nucl. Phys. {\bf A579},  596  (1994).

\bibitem{belt:70a}
Belt, B.~D., \emph{et al.}: Phys.\@ Rev.\@ Lett. {\bf 24},  1120  (1970).

\bibitem{vetterli:85a}
Vetterli, M.~C., \emph{et al.}: Phys.\@ Rev.\@Lett. {\bf 54},  1129  (1985).

\bibitem{pitts:88a}
Pitts, W.~K., \emph{et al.}: Phys.\@ Rev. {\bf C37},  1  (1988).

\bibitem{jourdan:85a}
Jourdan, J., \emph{et al.}: Phys. \@ Lett. {\bf 162B},  269  (1985).

\bibitem{schmid:96a}
Schmid, G.~J., \emph{et al.}: Phys.\@ Rev. {\bf C53},  35  (1996).

\bibitem{browne:96a}
Browne, K.~P., \emph{et al.}: Phys.\@ Rev. {\bf C54},  1538  (1996).

\bibitem{kosiek:66a}
Kosiek, R., M\"uller, D, Pfeiffer, R.: Phys.\@ Lett. {\bf 21},  199  (1966).

\bibitem{faul:80a}
Faul, D.~D., \emph{et al.}: Phys.\@ Rev.\@ Lett. {\bf 44},  129  (1980).

\bibitem{skopik:81a}
Skopik, D.~M., \emph{et al.}: Phys.\@ Rev. {\bf C24},  1791  (1981).

\end{thebibliography}

\end{document}